\documentclass[final]{ws-ijmpc-this}

\usepackage{graphicx}
\newcommand{\onlinecite}{\refcite}

\begin{document}

\markboth{George A. Baker and James P. Hague}
{From binary to continuous opinion dynamics}

\title{Rise of the centrist: from binary to continuous opinion dynamics}

\author{George A. Baker}
\address{Department of Physics, Loughborough University, Loughborough, LE11 3TU, UK}

\author{James P. Hague}
\address{Department of Physics, Loughborough University, Loughborough, LE11 3TU, UK}

\maketitle

\begin{history}
\received{31st March 2008}
\end{history}

\begin{abstract}
We propose a model that extends the binary ``united we stand, divided
we fall'' opinion dynamics of Sznajd-Weron to handle continuous and
multi-state discrete opinions. Disagreement dynamics are often ignored
in continuous extensions of the binary rules, so we make the most
symmetric continuum extension of the binary model that can treat the
consequences of agreement (debate) and disagreement (confrontation)
within a population of agents. We use the continuum extension as an
opportunity to develop rules for persistence of opinion
(memory). Rules governing the propagation of centrist views are also
examined. Monte Carlo simulations are carried out. We find that both
memory effects and the type of centrist significantly modify the
variance of average opinions in the large timescale limits of the
models. Finally, we describe the limit of applicability for Sznajd-Weron's
model of binary opinions as the continuum limit is approached. By
comparing Monte Carlo results and long time-step limits, we find that
the opinion dynamics of binary models are significantly different to
those where agents are permitted more than 3 opinions.

\end{abstract}


\ccode{PACS Nos.: 89.65.-s, 89.75.-k, 89.65.Ef}

\section{Introduction}

A recent revival of the study of social physics has been motivated by
the hope that qualitative understanding of the mechanisms of human
activity can be gained by constructing minimal models of interpersonal
interaction \cite{schelling1978a,ball2004a}. A number of `standard
models' have been introduced to describe diverse subjects such as
opinion dynamics, traffic flow, financial transactions, structure of
businesses, segregation and globalization. In this article we develop
a continuum extension to a model of binary opinion evolution.

We begin this article with a brief overview of approaches that have
been used to study opinion dynamics. Crucial to the current article,
we start with the binary `United we Stand, Divided we Fall' (USDF)
model (sometimes known as the Sznajd model) which is a lattice based
approach \cite{sznajdweron2000a}. The binary USDF model originated as
a modified Ising system, where up-spins map to opinion A and
down-spins to opinion B. A simple rule set is used to modify the
opinions of randomly selected pairs of agents on each time step. If a
set of neighbors share the same opinion, then their neighbors are
influenced and take on that opinion. If the neighbors have opposing
opinions, then their neighbors also take on opposing opinions. The
`community' is closed, meaning that there are no external influences
on the agents.

In the binary USDF model, relaxation times (time taken for average
opinion to change from a majority of As to a majority Bs) were found
to be in the region of 10000 time steps. It was observed that
avalanches of opinion changes could follow from single opinion
changes. There are three possible stable solutions at large
timestep. Either the whole population of agents takes the opinion A,
the entire population evolves to opinion B or an alternating opinion
(solution with modulated opinion, ...ABAB...) is reached \footnote{In
analogy to the ising model, some authors call this an
antiferromagnetic solution}. The relationship between the ratio of A
and B opinions at the start of the simulation and the large timestep
outcome was investigated, leading to the conclusion that an initial
concentration of 70\% A opinions is required to guarantee a final
state of all A. To simulate the effects of influences beyond a closed
community, `social noise' was introduced, where opinions change at
random with probability $p$ on each time step, leading to indefinite
opinion fluctuations. Stauffer {\it et al.} extended Sznajd-Weron's model to
represent opinion dynamics on a square lattice \cite{stauffer2000a}.

Using the Sznajd model as a basis, Fortunato developed a continuum
model \cite{fortunato2004a}. Since the Sznajd model has only two
opinions, it is of clear interest to develop a model with multiple
opinions. Fortunato developed a model for a continuous set of opinions
by using an agreement parameter to determine if agents interact
\cite{fortunato2004a}. Initially all agents are assigned random
numbers representing opinions on a continuum between $0$ and $1$. If
two neighbors' opinions agree to within this parameter then they are
considered `compatible' and a mean of their opinion is taken and
mapped onto the surrounding neighbors. The update rule in Fortunato's
model is similar to that of Deffuant {\it et al.} if Deffuant's
convergence parameter is set to $1/2$ \cite{deffuant2002a}.

The justification for an agreement parameter is that people tend to
form interest groups with others of similar viewpoint, and in that way
are in a better position to propagate their view. In Fortunato's model,
only those who are willing to discuss their opinion (i.e. agents that
have compatible opinions) are able to propagate their opinions. The
results of Fortunato are interesting; however, in our opinion the
model of Fortunato is not in the spirit of USDF models where agents of
strongly differing opinion have a symmetric influence on the opinions
of the population \footnote{In Fortunato's article, there is a brief
discussion of interactions between agents where opinions are outside
the tolerance of the agreement factor}. Thus, the polarizing effects
of confrontation are not fully considered. We return to this point in
section \ref{sec:continuum}. We discuss the detailed rules of the
binary USDF model and our extension to it in section
\ref{sec:binary}. For completeness, we also mention that alternative
binary and ternary models have been introduced by Galam
\cite{galam1990a} and Gekle {\it et al.}  \cite{gekle2005a}
respectively.

Network models of opinion dynamics have also been
constructed. Effective descriptions of the social networks within
populations have recently been developed
\cite{barbarasi1999a}. Stauffer and Meyer-Ortmanns proposed a model of
opinion dynamics which describes the interaction between agents on a
scale-free network \cite{stauffer2004a}, drawing on an earlier model
of Deffuant {\it et al} \cite{deffuant2002a}. Update rules are similar
to those used in Fortunato's extension of Sznajd-Weron's model
\cite{fortunato2004a}.

This article is set out as follows. In section \ref{sec:model}, we
describe the detailed rules of the binary USDF model (section
\ref{sec:binary}). Then in section \ref{sec:continuum}, we discuss
what we believe to be the most faithful continuum extension to
Sznajd-Weron's model, introducing a ruleset which leads to a true USDF model
with persistence of opinion (memory) and that handles centrists in a
consistent manner. We present results from Monte Carlo simulations of
the continuum model in section \ref{sec:continuumresults} and simulate
how the model evolves when the opinions of the agents are changed from
binary through multi-state to continuum in section
\ref{sec:discreteresults}. Finally, we summarize our results and
discuss possibilities for further extensions to the model in section
\ref{sec:conclusions}.

\section{Developing a continuum USDF model}

\label{sec:model}

\subsection{The binary USDF model}

\label{sec:binary}

The binary USDF Model \cite{sznajdweron2000a} is a modified Ising
model for simulating opinion dynamics where `spins' represent
opinions. The model community is set up as a set of agents $\{ S_i
\}$, each of which have ``opinions'' chosen independently and at
random from the discrete values $S^{0} \in \{ -1,+1 \}$. The agents
exist on a chain with periodic boundary conditions. In the following,
we use $S$ to represent binary opinions (the same notation as
Sznajd-Weron). Continuous opinions are represented by $O$. Once a
population has been set up, pairs of neighboring sites at lattice
points $i$ and $i+1$ are chosen at random on each timestep, $t$, and
the following rules are applied.

\begin{description}
\item[{\bf Rule S1}] Agreement: If a near-neighbor pair chosen at random has
the same opinions then the next-nearest neighbors take on those
opinions.
\begin{description}
\item[{\bf Condition S1}] If $S^{t}_i S^{t}_{i+1} = 1$ then apply Update S1.
\item[{\bf Update S1}] $S^{t+1}_{i-1} := S^{t}_{i}$ and $S^{t+1}_{i+2} := S^t_{i+1}$.
\end{description}
\item[{\bf Rule S2}] Disagreement: If a pair chosen at random has opposite
opinions then the surrounding neighbors take on opposing
opinions. 
\begin{description}
\item[{\bf Condition S2}] If $S^{t}_iS^{t}_{i+1} = -1$ then apply Update S2
\item[{\bf Update S2}] $S^{t+1}_{i-1}:=S^{t}_{i+1}$, $S^{t+1}_{i+2}:= S^{t}_{i}$
\end{description}
\end{description}

The model can be written with a single rule which is applied on each
timestep ($S^{t+1}_{i-1}:=S^{t}_{i+1}$, $S^{t+1}_{i+2}:=S^{t}_{i}$),
as was noted by Behera and Schweitzer \cite{behera2003a}. This
indicates that the model is trivial in the large timestep limit, since
the lattice can be separated into two sublattices, with a
ferromagnetic configuration always forming on a single
sublattice. Therefore there are 3 outcomes: (a) Both sublattices have
average opinion, $S_{\rm av} = -1$ (b) Sublattices have opposite
average opinions - there are two configurations with this type of
order, and thus $S_{\rm av}=0$ (c) Both sublattices have $S_{\rm av} =
1$. While Sznajd-Weron's model may be trivial in the large timestep
limit, we believe that the ``united we stand, divided we fall''
sentiment of the model is interesting, and have aimed to properly
include the basic idea in a continuum extension.

\subsection{Developing an extended model of continuous opinions}

\label{sec:continuum}

A key factor of the binary USDF model is that it is not only agents
with similar opinions that interact and propagate their opinions;
agents with diametrically opposed opinions may also interact. Our
motivation is to develop a model with continuous opinions with the
same core idea. We use the binary USDF consensus model
\cite{sznajdweron2000a} as a base from which to develop, and introduce
a set of rules modified from those of a previous continuum extension
developed by Fortunato, which only included propagation of similar
opinions \cite{fortunato2004a}.

In the model of Fortunato \cite{fortunato2004a}, an agreement is defined if two
opinions are within a set margin, i.e. $|O^{t}_i - O^{t}_{i+1}|\le\eta$. (In
some regards, this is quite similar to Axelrod's cultural propagation
model \cite{axelrod1997a}). We also want to include both the continuum
nature of Fortunato's model and an interaction between strongly
conflicting agents similar to that in Sznajd-Weron's USDF model. The most
symmetric way of doing this is to define a disagreement if $|O^{t}_i +
O^{t}_{i+1}|\le\eta$, where $\eta$ is a tolerance defining when opinions
are of similar magnitude (we call $\eta$ the agreement parameter). It
is easy to see that if neighboring opinions are nearly exactly
opposite, this condition holds.

In Ref. \onlinecite{fortunato2004a}, only agreement is treated in this
way. There is a brief discussion of the disagreement required to
provoke a ``divided we fall'' response in
Ref. \onlinecite{fortunato2004a}, but where that was implemented in
Ref. \onlinecite{fortunato2004a}'s model it was assumed that any pairs of
agents that do not agree, disagree, and the neighboring agents then
take on the staggered alternating opinions of the pair of agents (this
is equivalent to always making the map $O^{t+1}_{i-1}:=O^{t}_{i+1}$
and $O^{t+1}_{i+2}:=O^{t}_{i}$ without a test). We do not feel that
this is in the spirit of the the binary USDF model, since there is not
symmetry between strong disagreement and strong agreement. We
introduce disagreement in the most symmetric extension to the
agreement rule in {\bf Rule 2}, which is one of the main differences
between our model and Fortunato's (there are also other differences as
discussed in this section).

\begin{description}
\item[Rule 1] (Agreement) If two agents have opinions that agree to
within a certain margin, then they are said to agree. Then their
average opinion propagates to neighboring sites.
\begin{description}
\item[Condition 1] If $|O^{t}_i - O^{t}_{i+1}|\le\eta$ then update opinions
with update 1.
\item[Update 1]
$O_{i+2}^{t+1} := O_{i-1}^{t+1}:= (O_{i}^{t} + O_{i+1}^{t})/2$
\end{description}
\item[Rule 2] (Disagreement) If two agents have opinions that are opposite
to within a certain margin, then they are said to disagree. Then their
difference in opinion propagates to neighboring sites.
\begin{description}
\item[Condition 2] If $|O^{t}_i + O^{t}_{i+1}|\le\eta$ then update opinions
with update 2.
\item[Update 2]
$-O_{i+2}^{t+1} := O_{i-1}^{t+1}:= (O_{i+1}^{t} - O_{i}^{t})/2$
\end{description}
\end{description}
Therefore, there is a slight additional bias in the model towards the
propagation of groups of agents with strongly opposed opinions. A
motivation for this rule is that if there are strong disagreements
between groups with diametrically opposed opinions, other agents are
polarized into opposing viewpoints. There may be examples of such
spreads (for instance in clashes between communists and fascists in
inter-war Europe, or during conflicts between tribal groups where a
positive feedback mechanism causes small frictions to turn into large
disagreements) although we prefer to leave interpretation of the rule
set to social scientists.

We note that the binary USDF model is not able to take into account
the effects of intrinsic persistence of opinion (we call this
memory)\footnote{We note that the phrase `persistence of opinion' is
used to denote an extrinsic quantity relating to the length of time
between opinion flips in some articles \cite{stauffer2002a}}. Using a
continuum model is an excellent opportunity to include persistence
effects. Memory is a strong characteristic of opinion dynamics and we
believe that at least an investigation of its effects on the model is
important. Using a memory factor means that the way an opinion is
modified between timesteps is determined from both the initial opinion
of the agent and updates 1 and 2 of the continuum USDF
model. Therefore an agent's opinion does not switch from one end of
the opinion scale to the other instantly; rather the agent requires
persuasion over a number of timesteps. The simplest way of
implementing memory is to take a linear combination of the existing
opinion and the opinion determined from updates 1 or 2. This can only
be implemented in the continuous version of the USDF model, and not in
the binary version (since the value of $S=1/2$ does not exist in the
set of opinions used in the binary model), showing one of the
advantages of our implementation.

To introduce a memory factor we make the following modification to updates
1 and 2,
\begin{description}
\item[Rule M1] Agreement rule modified for persistence of opinion
\begin{description}
\item[Update M1a]
$O_{i-1}^{t+1}:= \alpha O_{i-1}^{t} + (1-\alpha) ( O^{t}_{i+1} + O_{i}^{t} )/2$
\item[Update M1b]
$O_{i+2}^{t+1}:= \alpha O_{i+2}^{t} - (1-\alpha) ( O^{t}_{i+1} + O_{i}^{t} )/2$
\end{description}
\item[Rule M2] Disagreement rule modified for persistence of opinion
\begin{description}
\item[Update M2a]
$O_{i-1}^{t+1}:= \alpha O_{i-1}^{t} + (1-\alpha) ( O_{i+1}^{t} - O_{i}^{t} )/2$
\item[Update M2b]
$O_{i+2}^{t+1}:= \alpha O_{i+2}^{t} - (1-\alpha) ( O_{i+1}^{t} - O_{i}^{t} )/2$
\end{description}
\end{description}
such that in a system where all agents have opinions $O$
with same magnitude, the total strength of opinion does not grow or
shrink spontaneously. The index $t$ in the $O^{t}_n$ notation relates
to the generation (time step) of opinions. $\alpha$ denotes how much
relative influence memory or interaction with neighboring agents has
on the new generation of opinions. We summarize the pairwise
interaction in figure \ref{fig:schematic}.

\subsection{An Ambiguous Region}

\begin{figure}
\includegraphics[width=120mm]{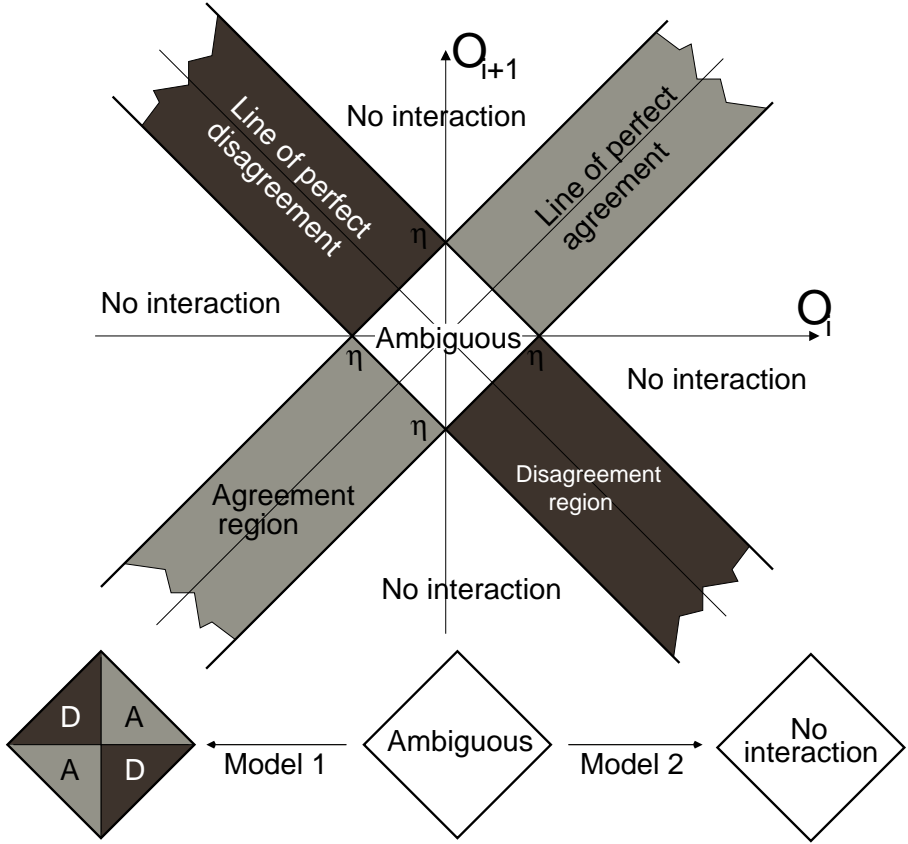}
\caption{Schematic showing the interaction between neighboring
agents. When two near neighbors have the same strength of opinion to
within a tolerance $\eta$ (i.e. $|O_{i} - O_{i+1}|<\eta$), they are
said to agree. When the strength of opinion is opposite to within
$\eta$ (i.e. $|O_i + O_{i+1}|<\eta$), they are said to disagree. For
all other cases the neighbors are non-interacting. There is an area
of ambiguity associated with these definitions for small magnitude (or
centrist) opinions where $|O_i|+|O_{i+1}|<\eta$. There are two
resolutions to the ambiguity. In the lazy centrist model, neighbors
are be considered to be non-interacting if $|O_i|+|O_{i+1}|<\eta$
(ambivalent to each other). In the active centrist model, the type of
interaction between centrist neighbors is determined via a modified
Sznajd USDF rule (agree if $|O^{t}_i|+|O^{t}_{i+1}|<\eta$ and
$O^{t}_{i}O^{t}_{i+1}>0$ and disagree if
$|O^{t}_i|+|O^{t}_{i+1}|<\eta$ and $O^{t}_{i}O^{t}_{i+1}<0$). The
second variant is mathematically equivalent to the binary USDF model
when $\eta\rightarrow\infty$ if $O_i/|O_i| \mapsto S_i $. Both
variants are identical to the binary USDF model if the agents are
initialized with binary opinions $O^{0}_{i} \in \{ +1 , -1\}$, no memory
parameter is used ($\alpha = 0$) and $\eta < 2$.}
\label{fig:schematic}
\end{figure}

The new rule set has an area of ambiguity for opinions of weak
magnitude, since there is a small region of the interaction where both
conditions 1 and 2 apply. The opinions that lie within this region are
all smaller in magnitude than the agreement parameter and occur when
$|O_{i+1}|+|O_{i}|<\eta$.  In this way, they can be thought of as
representing centrist opinions. This can clearly be seen in figure
\ref{fig:schematic}; within the central region of the graph. We treat
this region separately. There are two ways to deal with the low
opinion region (a) interacting centrists (active centrist model) and
non-interacting centrists (lazy centrist model).

\subsubsection{Active centrist model (ACM)}

A simple way to remove the ambiguity that arises for weak opinion is to treat separately the area
where both conditions 1 and 2 are satisfied. Our first option for determining
if agreement or disagreement rules should be applied is in keeping
with the binary USDF Model \cite{sznajdweron2000a}. We split the low
opinion region into quadrants, so that the sign of the product of the
opinions determines the type of update. A positive product is defined
as agreement, and a negative product is defined as disagreement.
\begin{description}
\item[Rule A] Additional rule for ambiguous region of ACM when
$|O^{t}_{i+1}|+|O^{t}_{i}|<\eta$.
\begin{description}
\item[{\bf Condition A1:}] There is agreement if $O^{t}_{i}O^{t}_{i+1} > 0$ so
use update M1
\item[{\bf Condition A2:}] There is disagreement if $O^{t}_{i}O^{t}_{i+1} < 0$
so use update M2
\end{description}
\end{description}
We call this set of rules the active centrist model (ACM). There is a
direct mapping to the binary USDF Consensus Model when
$\eta\rightarrow\infty$ if we make the substitution $O_i/|O_i| \mapsto
S_i$.

\subsubsection{Lazy centrist model (LCM)}

An alternative way of treating weak opinions is to make no update if
an opinion pair falls within the ambiguous region. We call this
disambiguation the `lazy centrist model' (LCM). The evolution of this
model may be of interest to study disinterested voters. We note that
both ``active centrist'' and ``lazy centrist'' variants are identical
to the binary USDF model if the population of agents is initialized
with only $O_{i} \in \{-1,+1\}$, $\alpha = 0$ and $\eta < 2$.

\subsection{Algorithm and implementation}


To simulate the model, we use a simple Monte Carlo algorithm to
investigate how a population of opinions evolves over a period of
time. The program was written in Mathworks MATLAB which offers a long
period random number generator of the order $2^{1492}$, and a number
of nice prepackaged routines for analyzing the resultant
configurations of agents (e.g. convenient histogram routines).

Each run of the simulation is initialized on a 1D lattice, with each
site allocated a random opinion from $-1$ to $+1$. For the continuum
model, opinions are allocated with a uniform random variate. For the
discrete-opinion model, the opinions are initialized with
\begin{equation}
O^{0}_{i} := -1 + 2 m / (M - 1)
\label{eqn:discrete}
\end{equation}
where $M$ is the total number of opinions, and $m$ is a
uniform integer variate with $M$ levels $m\in \{ 0,\cdots,M-1\}$
(parameters of $\eta < 2/(M-1)$ and $\alpha=0$ ensure that only opinions
of the same magnitude interact). There is a choice of boundary conditions. Our population of opinions exists on a ring
(i.e. 1D chain with periodic boundary conditions). Considering that
the model that we propose is highly simplified, the choice of boundary
conditions is expected to be unimportant in comparison to other
simplifications (e.g. lattice vs network).

After the initial setup of the population, we simulate using a Monte
Carlo method. Our simulation uses sequential rather than concurrent
updating, where a pair of the population is selected at random on each
time step. The algorithm finishes when a certain preset number of time
steps have been completed. When we calculate ensemble averages,
typically we simulate 10000 ensembles each with a different random starting
population.

The algorithm can be summarized as follows:
\begin{enumerate}
\item Initialize the population with random opinions (spins), either
with continuous or discrete opinions.
\label{step:one}
\item Choose a pair of neighboring spins from the total population
\label{step:two}
\item Determine the required update step:
\label{step:three}
\begin{enumerate}
\item If the opinion pair is in the ambiguous region (condition A is
satisfied), update next-neighbor opinions according to Rule A
(ACM) or do nothing (LCM).
\item Otherwise, update using rule 1 or rule 2.
\end{enumerate}
\item Repeat steps \ref{step:two} and \ref{step:three} until the
required number of time steps have been simulated
\item Restart at step \ref{step:one} until the ensemble of runs is
large enough for meaningful statistical averages.
\end{enumerate}

\section{Results and Discussion}

\subsection{Properties of the continuum model}

\label{sec:continuumresults}

\subsubsection{Evolution of opinions}

\begin{figure}
\includegraphics[height=13cm,angle=270]{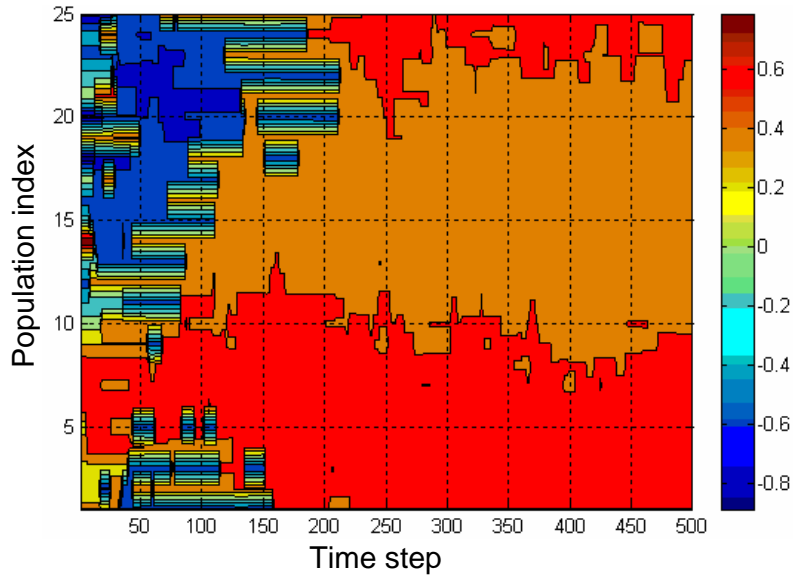}
\caption{(Color online) Figure showing opinion evolution of a
population of agents following the LCM rules. Here $\eta = 1$, $N=25$,
$T=500$, $\alpha = 0$. Towards the start of the simulation, the
population is multi-valued, with regions of modulated opinion and an
average negative opinion. Two domains of positive opinion form towards
the end of the simulation. We believe that this formation of domains
is similar to that in Axelrod's cultural model.}
\label{fig:three}
\end{figure}

We demonstrate the inner workings of our ``society'' of agents by
following the evolution of all the individual opinions in a single
run, shown in figure \ref{fig:three} as a density plot of the opinion
of each agent vs time step. A small population of $N=25$ agents is
simulated, using the LCM ruleset with agreement parameter $\eta=1$, no
persistence of opinion ($\alpha = 0$) over a period of 500 timesteps.
Each location on the $y$-axis represents an individual agent and time
progresses along the $x$-axis. The $z$-axis (color) represents
opinion. As expected, opinions are diverse at the beginning of the
simulation, since they have been initialized randomly. Over the course
of the simulation, patches of monotonous opinions become
dominant. Towards the end of the run shown in the figure, the opinions
converge to two regions of similar positive opinion. Since the
opinions are similar enough to continue interacting, it is expected
that a single opinion will eventually emerge. We believe the origins
of opinion domains to be similar to those in Axelrod's model of
cultural evolution \cite{axelrod1997a}. The regions of modulated
opinion that can be seen early in the simulation become extinct by
just over 200 time steps. This coincides with a flip from negative to
positive opinion values. Before timestep, $\tau=50$, a greater
proportion of society is at the `blue' end of the opinion
scale. Within a small number of time steps, the number of red opinions
increases dramatically. By $\tau = 200$, the `blue' opinions are
almost eradicated. This characteristic behavior is also seen in the
binary USDF model \cite{sznajdweron2000a}.

\subsubsection{Average opinion}

\begin{figure}
\includegraphics[height=13cm,angle=270]{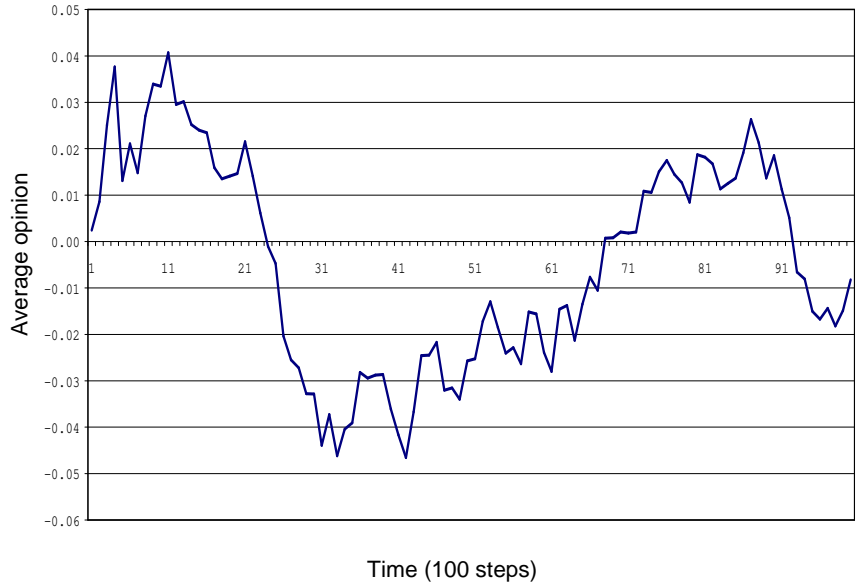}
\caption{A demonstration of what happens to the average opinion over
long time period during a single run. $N=250$ agents, $T_{max}=10000$
steps, $\eta=1$, $\alpha=0$ and the ambiguous region is treated using
the ACM rules.}
\label{fig:four}
\end{figure}

Figure \ref{fig:four} shows the evolution of the average opinion of a different population of agents during 
a single run over 10000 loops. The population size was $N = 250$ agents with
an agreement parameter of $\eta = 1$ and no memory effects
($\alpha = 0$). The ACM rule set was used. Only a very small change
in the average opinion was seen, however it is interesting to note that there were several intersections with the $x$-axis during the simulation. The swing from positive to negative opinion took place over about 4000 time steps. It is
important to remember that results obtained from single runs are
not predictive.

\subsubsection{Distribution of average opinions}

\begin{figure}
\includegraphics[height=62mm,angle=270]{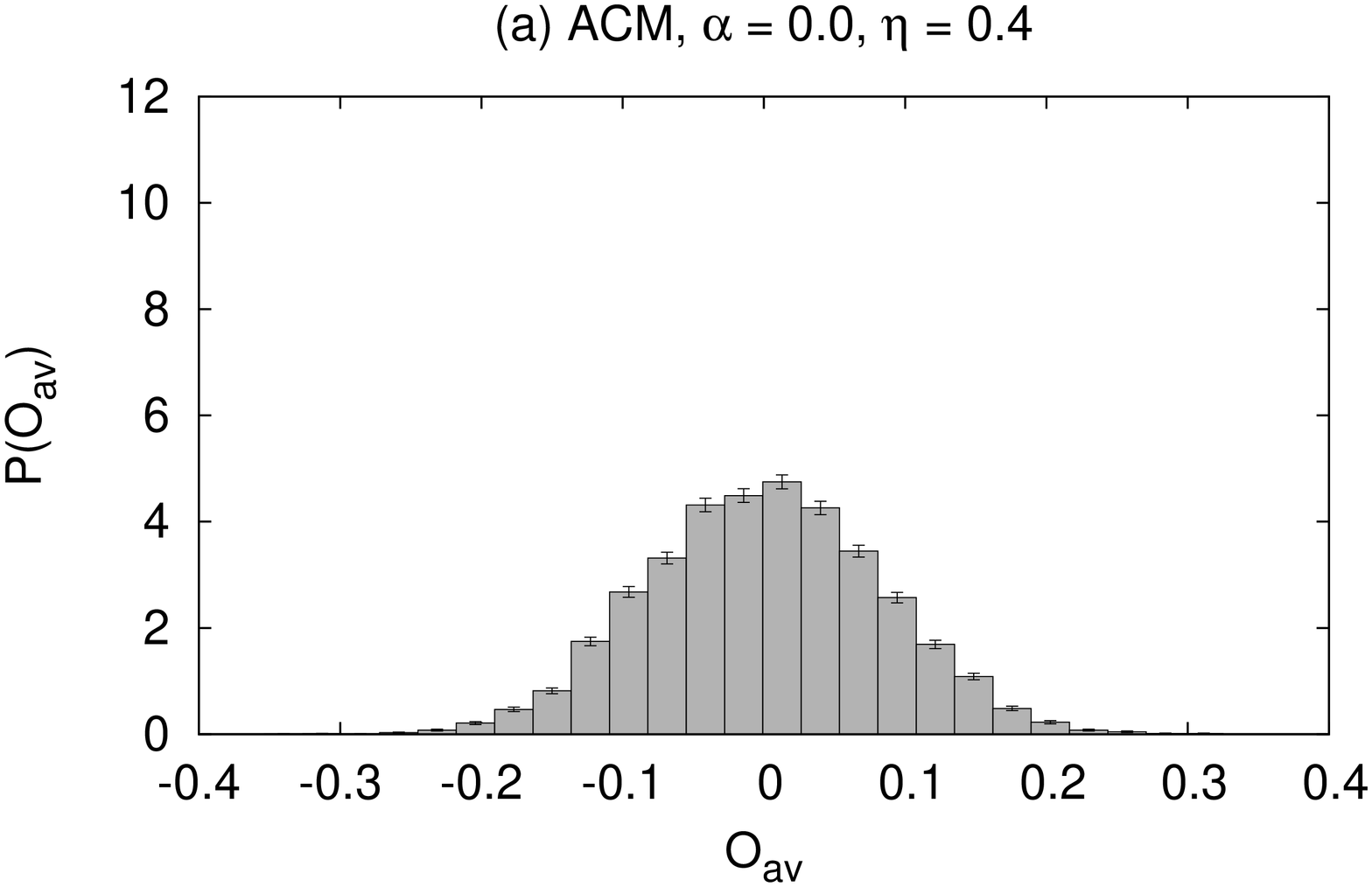}
\includegraphics[height=62mm,angle=270]{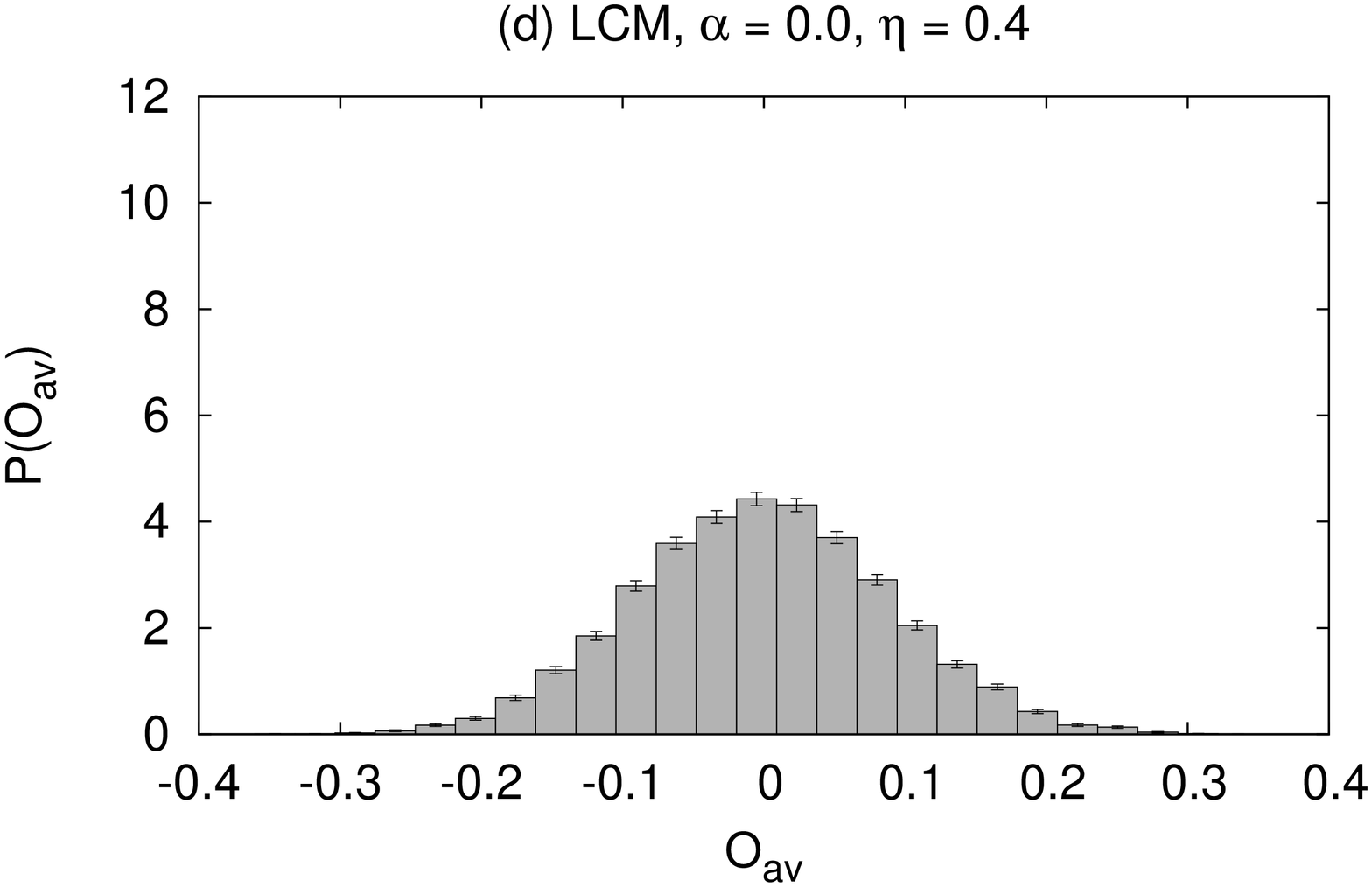}
\includegraphics[height=62mm,angle=270]{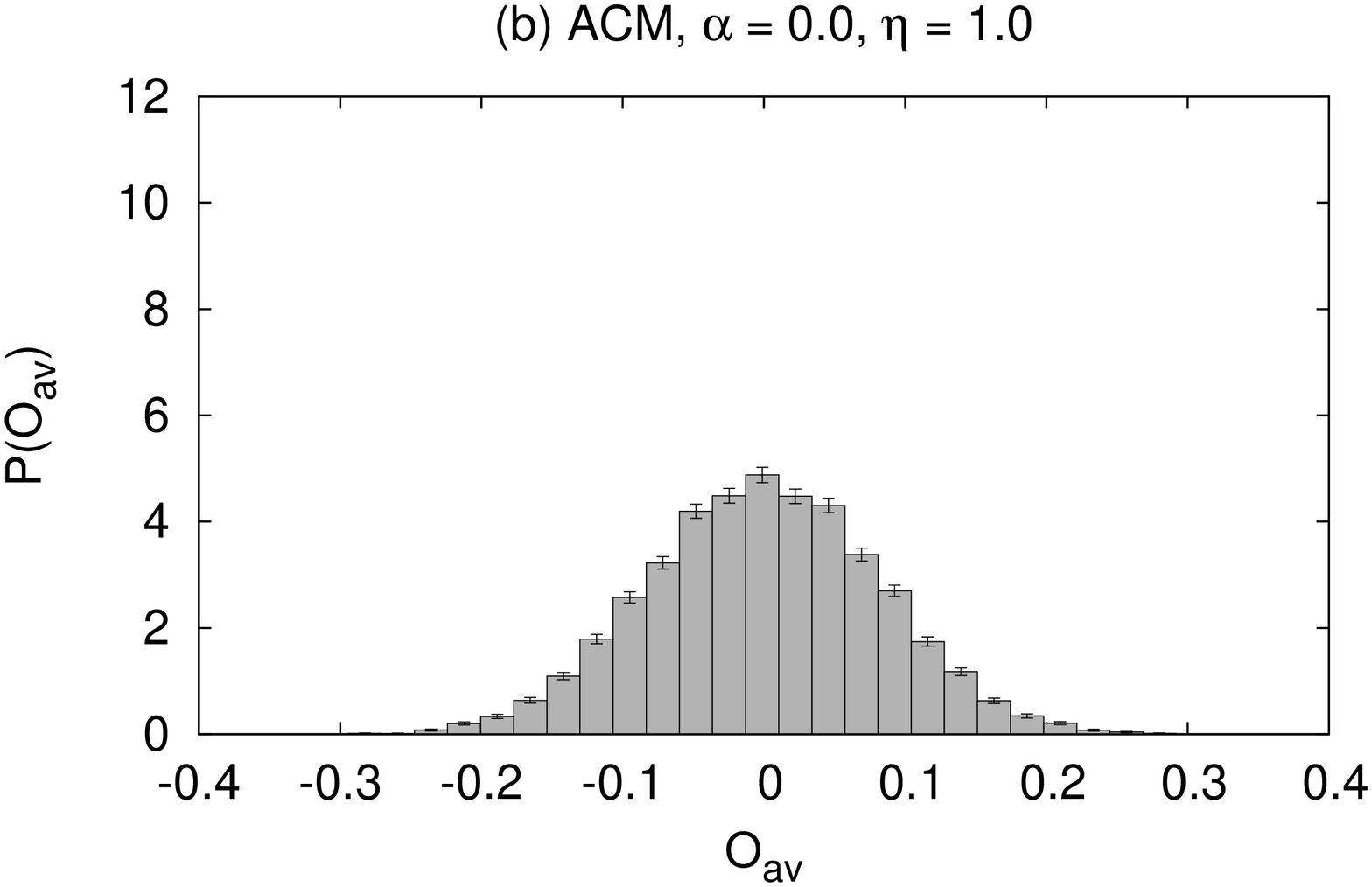}
\includegraphics[height=62mm,angle=270]{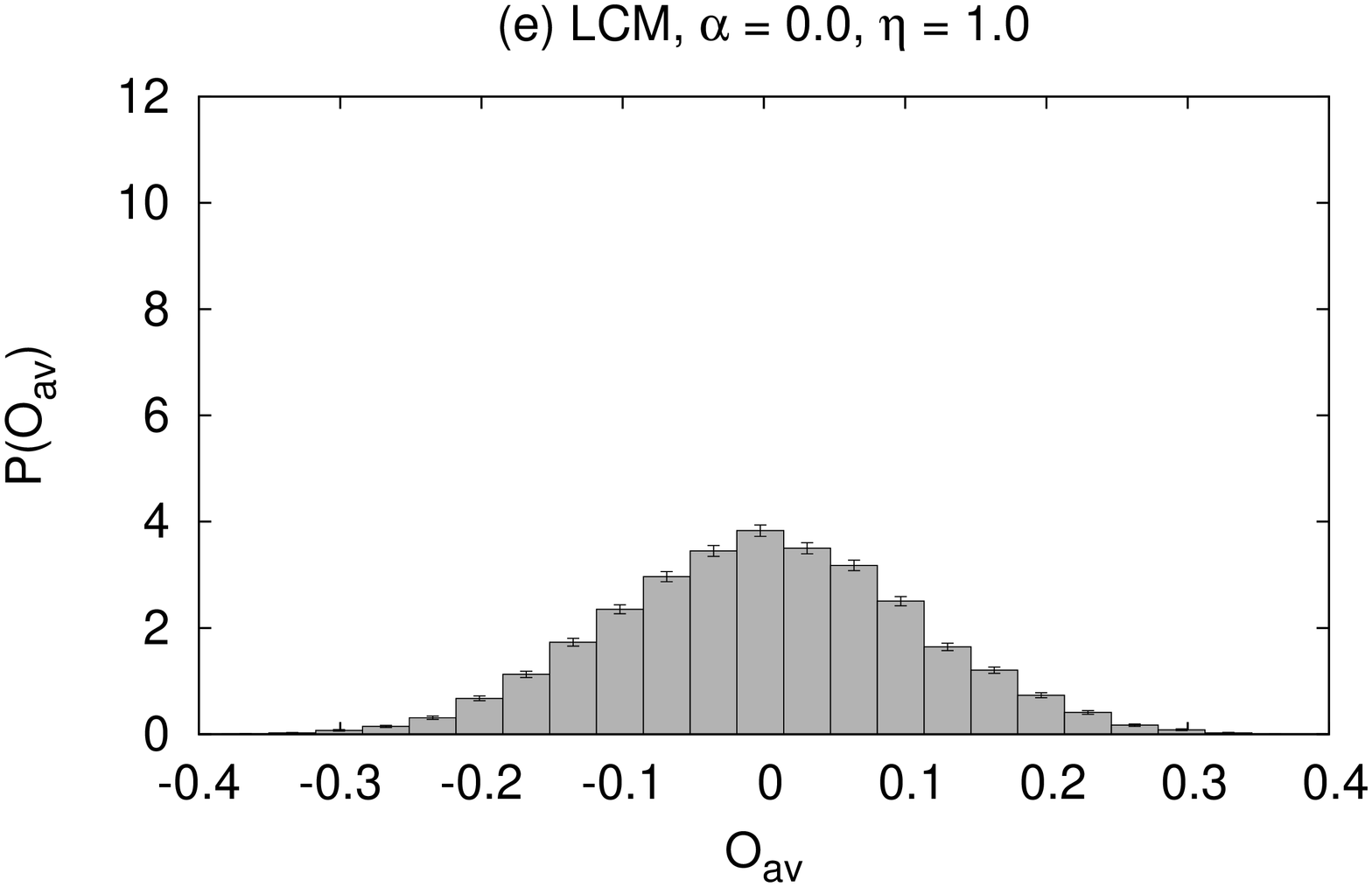}
\includegraphics[height=62mm,angle=270]{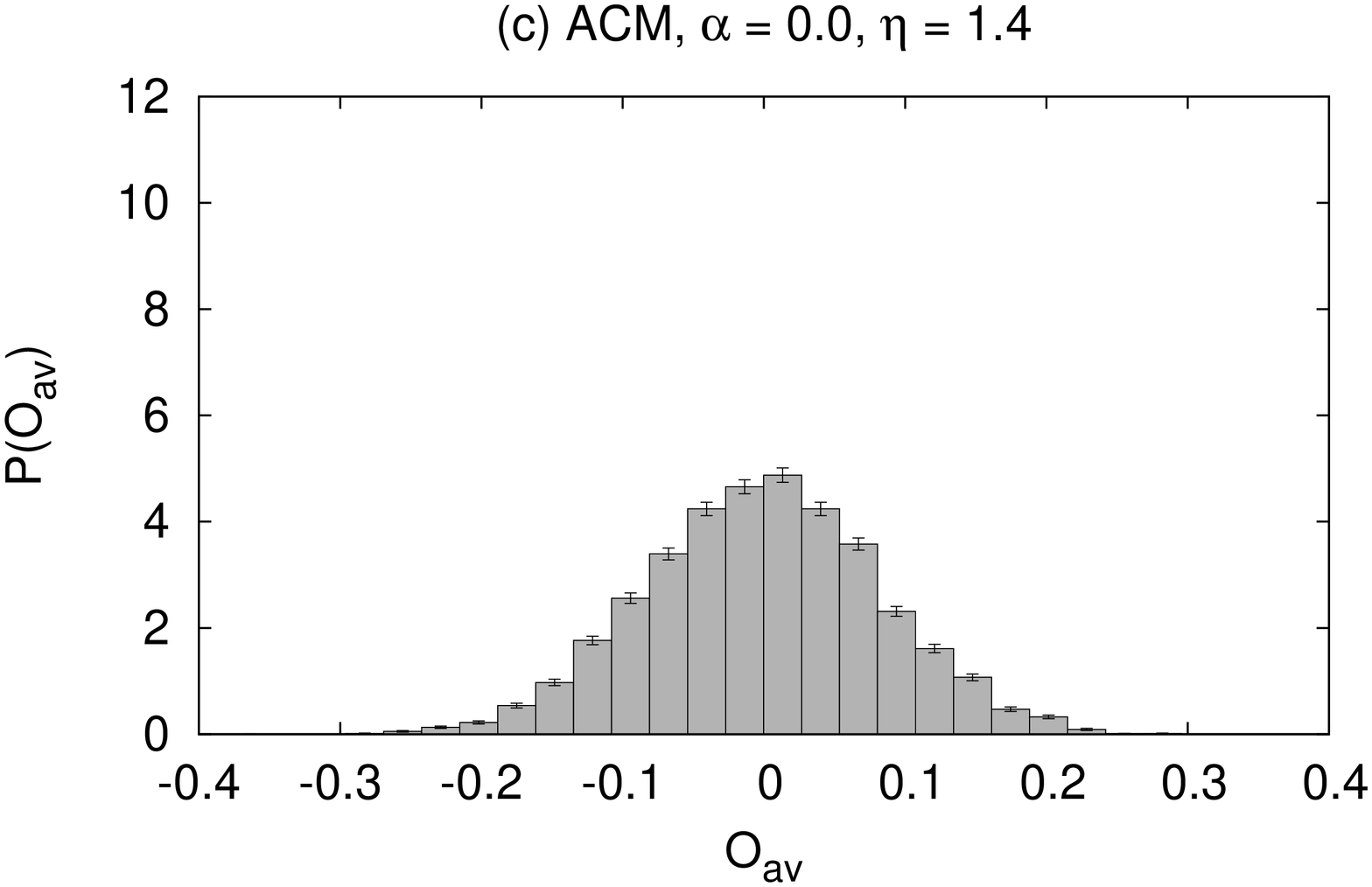}
\includegraphics[height=62mm,angle=270]{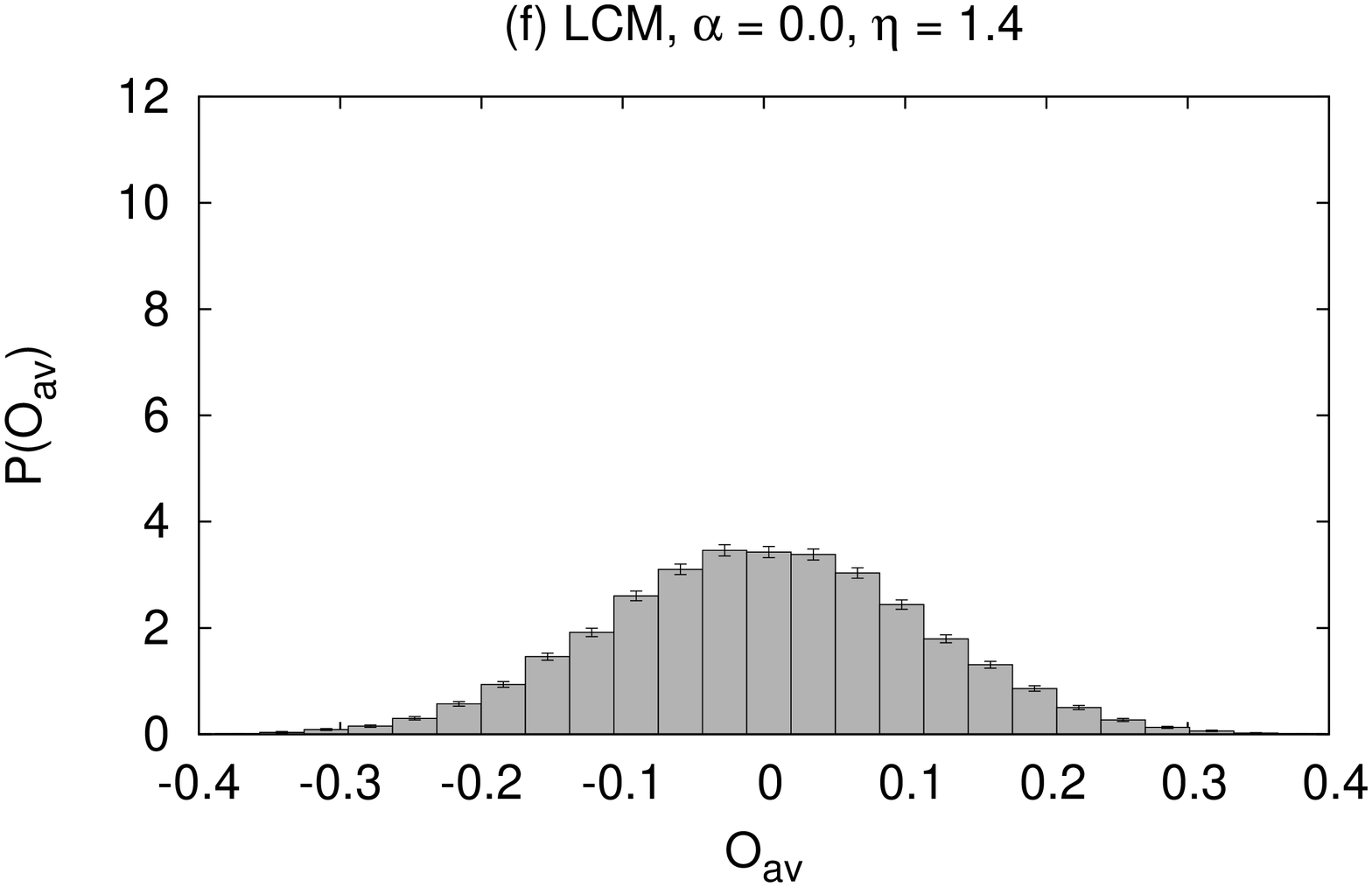}
\caption{Histograms showing the distribution of average opinions in an
ensemble of $N_{\rm ens} = 10000$ runs at timestep $T_{\rm max} = 10000$. The
plots are normalized so that the total area of the boxes is unity and
the graphs represent the density of average opinions. $N = 250$, $T_{max} = 10000$, $\alpha = 0$, $\eta = 0.4$. In general, the variance of the
LCM is higher. We discuss this point with regard to a 3 state model
later on in the article.}
\label{fig:five}
\end{figure}

\begin{figure}
\includegraphics[height=62mm,angle=270]{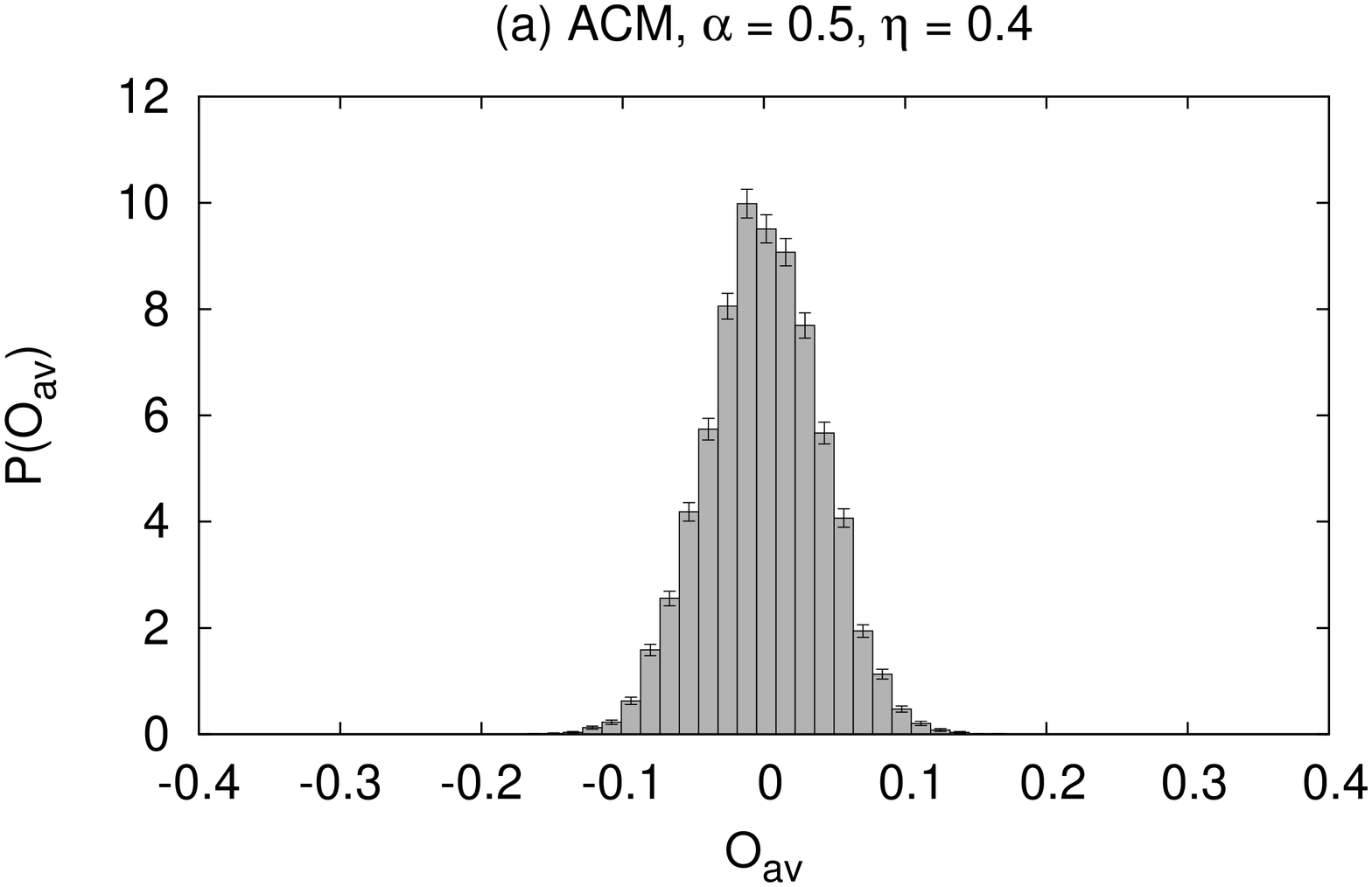}
\includegraphics[height=62mm,angle=270]{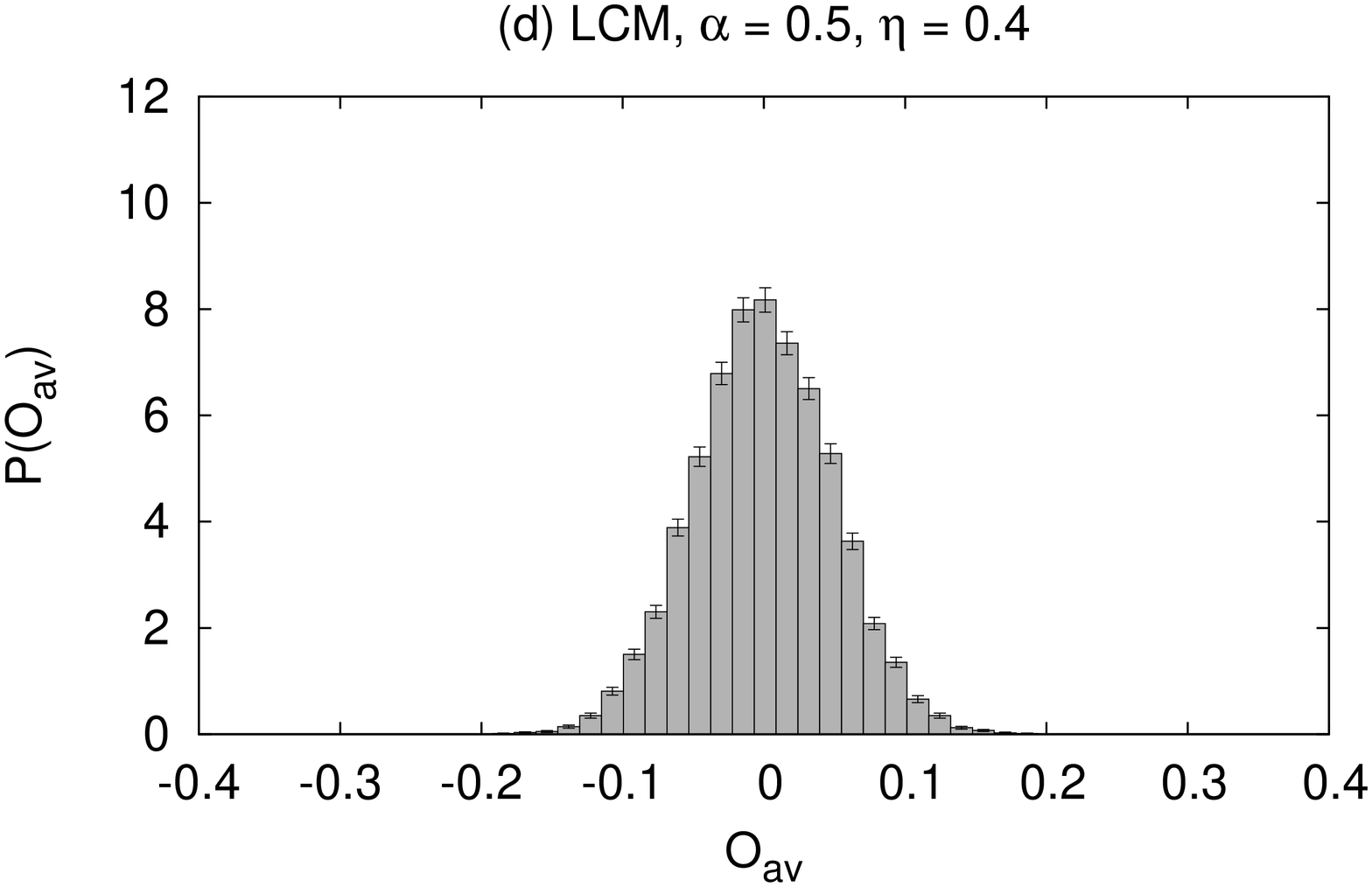}
\includegraphics[height=62mm,angle=270]{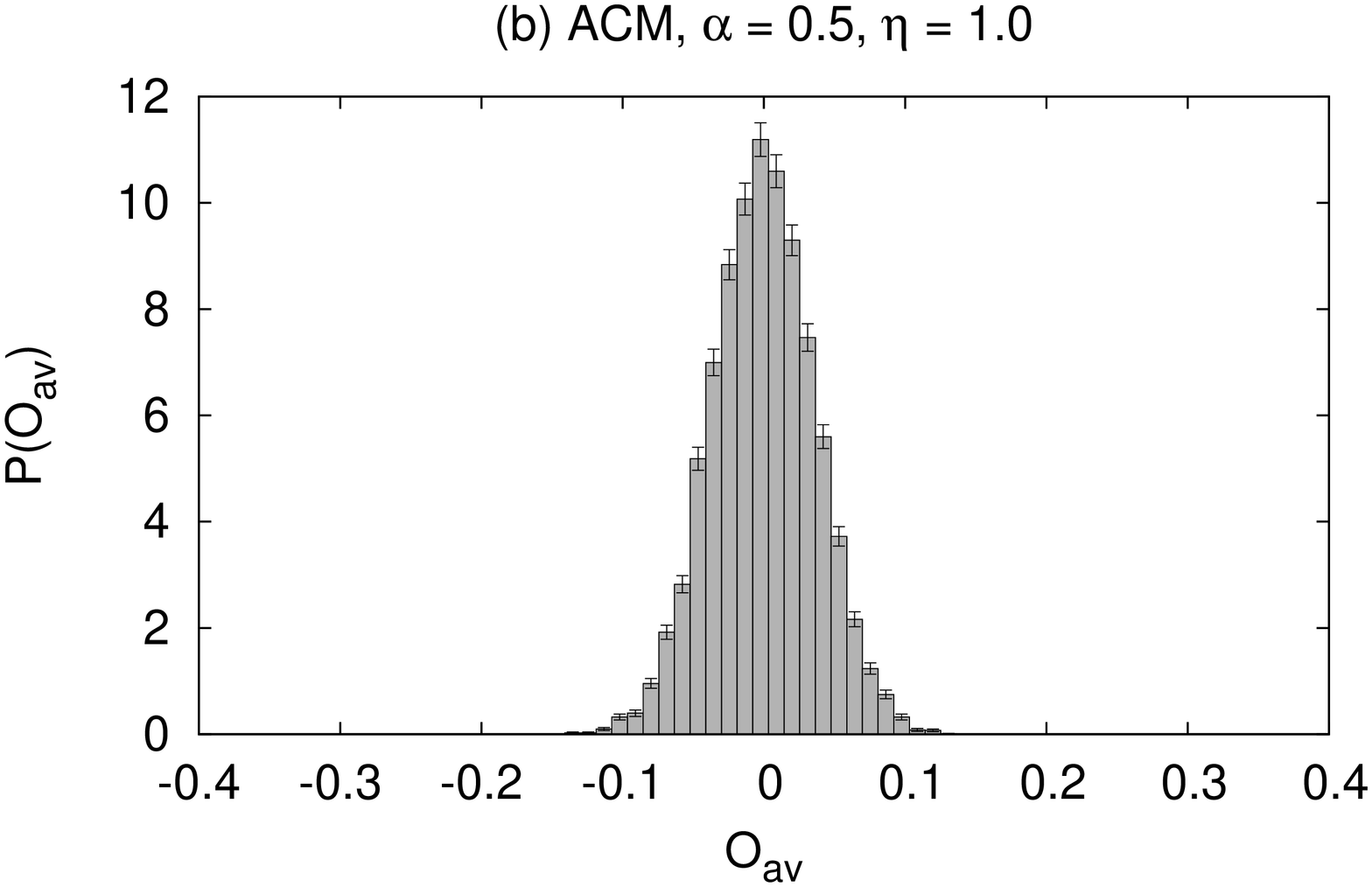}
\includegraphics[height=62mm,angle=270]{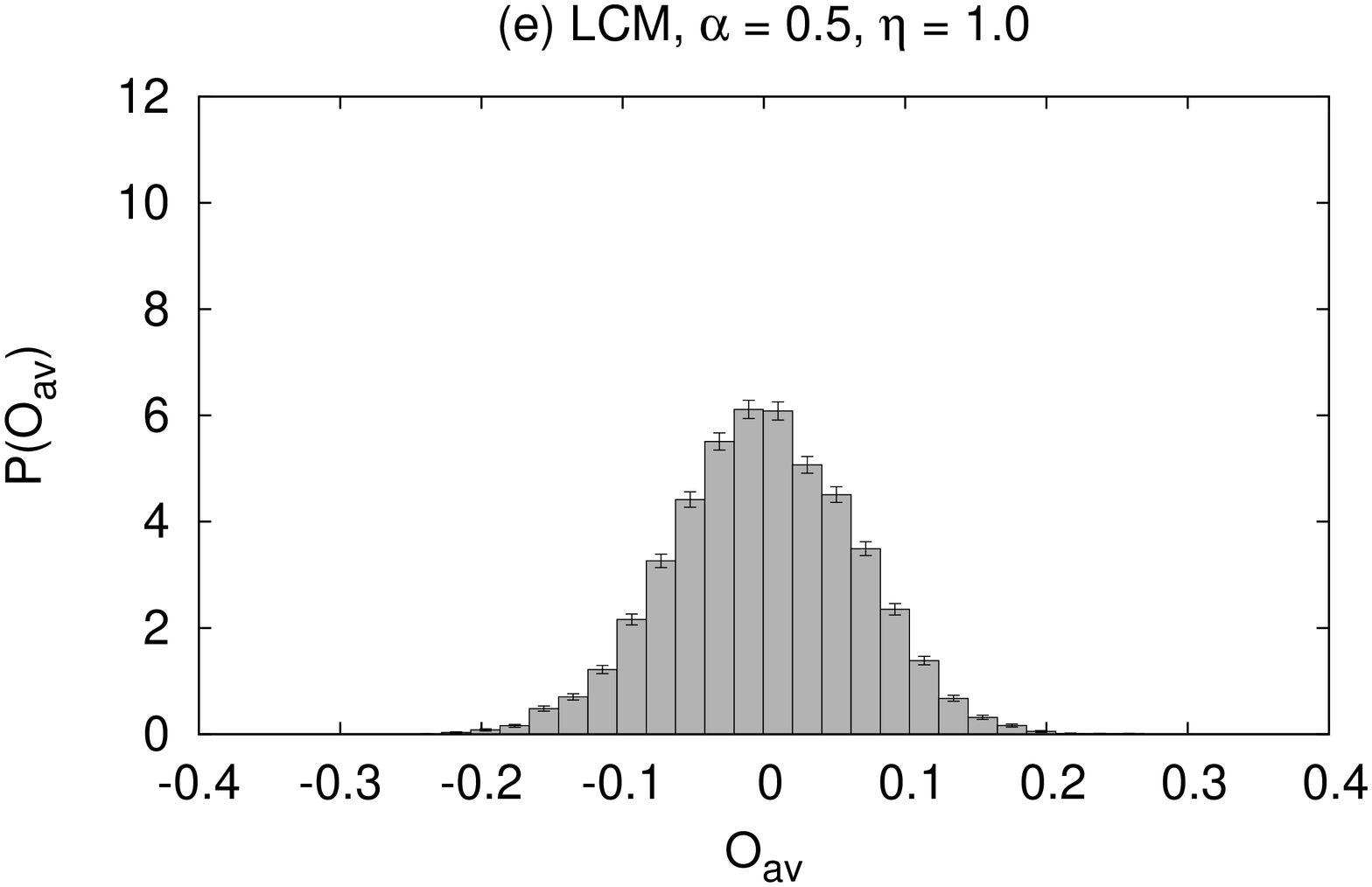}
\includegraphics[height=62mm,angle=270]{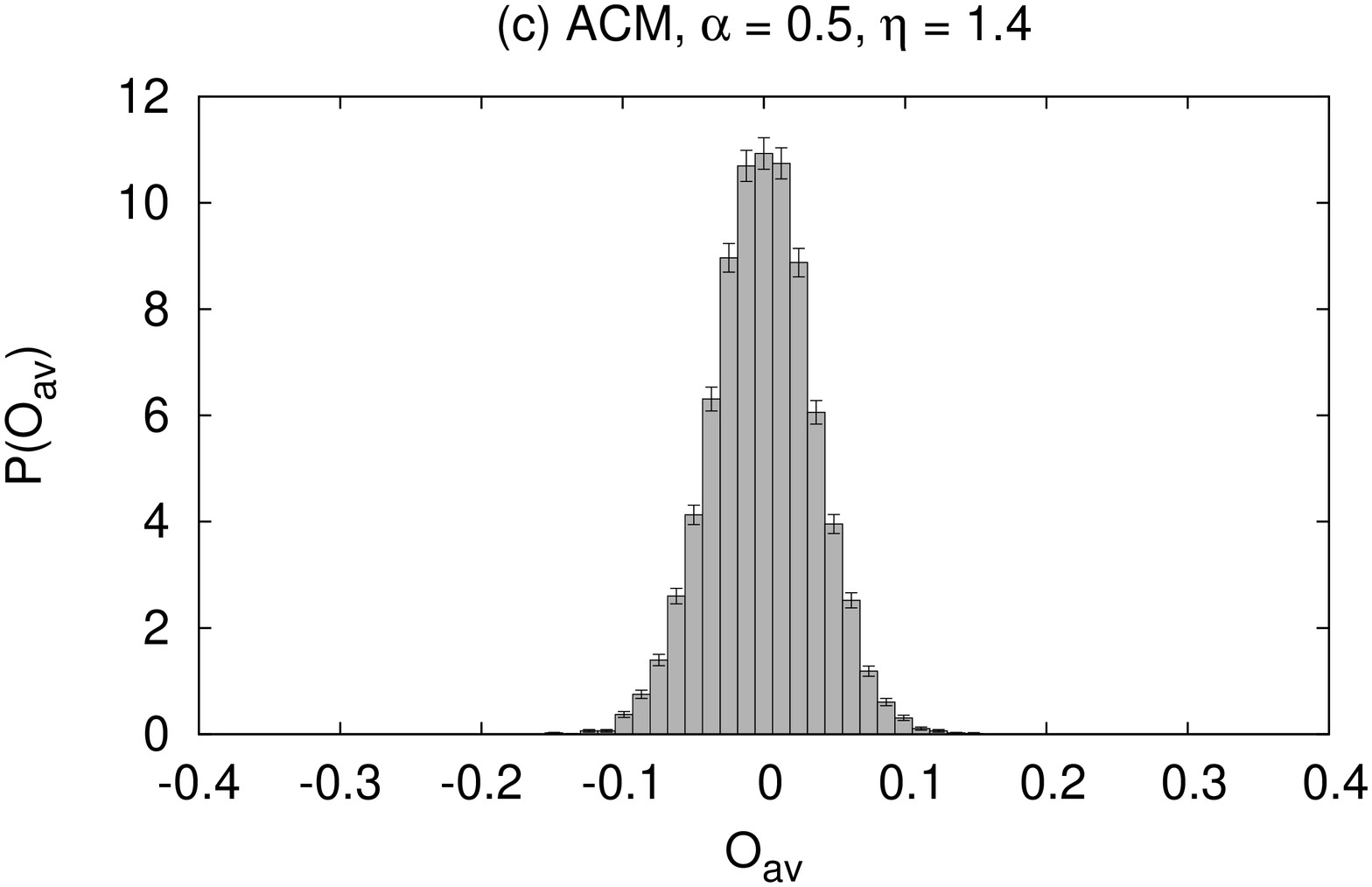}
\includegraphics[height=62mm,angle=270]{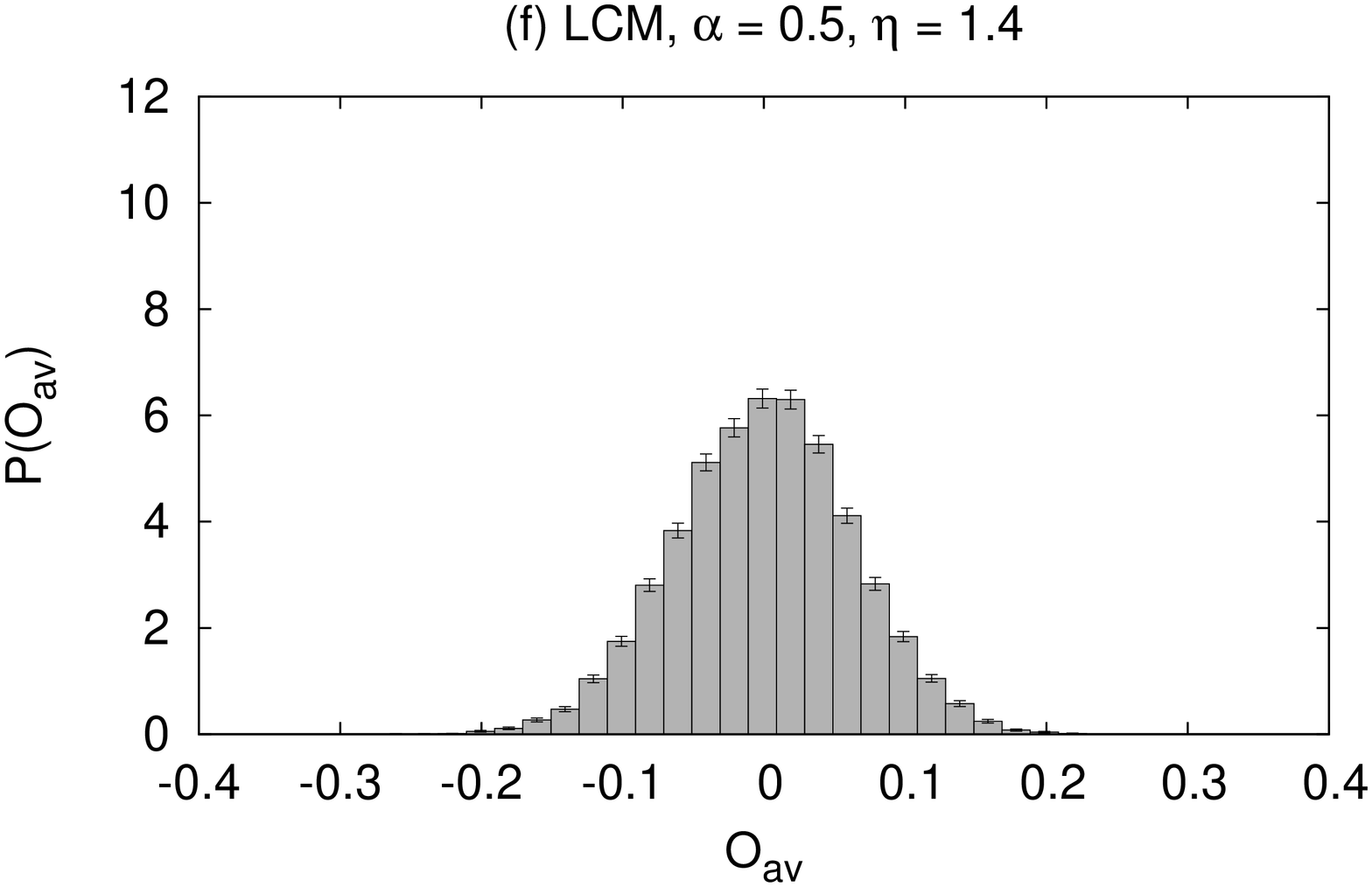}
\caption{As figure \ref{fig:five}, with persistence of memory $\alpha
= 0.5$, $N = 250$, $T_{max} = 10000$, $\alpha =
0.5$, $\eta = 0.4$. Again, the variance of the LCM is
higher. Persistence of memory makes the average opinion more likely to
be centrist as it draws opinions slowly to the average. LCM is
unlikely to have a steady state, since centrists do not propagate
their opinions, and are constantly ``dying out'' and being
replaced. The variance in the ACM barely changes.}
\label{fig:fivemem}
\end{figure}

In this section, we analyze the distribution of average opinions. After a designated number of time steps, the average opinion of a
single run is calculated. After 10000 ensemble runs, these averages are binned into a histogram. We use histograms to analyze the ensemble since it gives more information than the computation of a simple average or standard deviation. For instance, non-Gaussian distributions can be assessed. We show histograms computed for simulations with no memory factor ($\alpha = 0$) in figure \ref{fig:five}. The effects of the memory factor on the distribution are shown in figure \ref{fig:fivemem}.

The results in figure \ref{fig:five} can be directly compared with the binary USDF model. The binary USDF model \cite{sznajdweron2000a} has
three long time steady states, so if the continuum extension to the model has no effect, we would expect the distribution of average opinions
computed after a large number of time steps to be 3 peaked
(we confirm this later in the article). Clearly, our model has quite a
different distribution of opinions, with a broad single peaked
structure. In the LCM, increase in $\eta$ broadens the distribution. In particular, panel (f) shows that the probability distribution may be spreading into 2 peaks. The spread in the distribution is expected in the LCM rules, since agents with $O<\eta$ may not propagate their opinion, and so become extinct in the large time step limit. In contrast, little change is seen in the ACM as $\eta$ is changed. The different behavior of our model could be a result of several
factors. Either a steady state has not been reached, or the binary USDF Consensus Model breaks down as a set of continuous opinions are reached, with the absence of extreme opinion
sets. We revisit this point later in the article.

Figure \ref{fig:fivemem} shows the effect of memory. In both the LCM
and ACM models when a memory factor of $\alpha=0.5$ is used, the
distribution of average opinions is halved. In the limiting case of
$\alpha = 1$, the distribution of opinions may not change. This
indicates that one of the effects of memory is to increase the time
scale in the model. A secondary effect of memory on the LCM is
non-trivial. Consider a pair of agents that agree with positive
opinion $O=0.75$ sitting next to an agent with $O_{\rm nn} =
-0.75$. The effect of the influence of the first pair is to change the
opinion of the $O_{\rm nn}=-0.75$ agent to $O_{\rm nn}=0$. Thus,
persistence of opinion causes new opinions to be created, which can
have value $O<\eta$, so there is a constant flux between creation and
extinction of centrist views in the LCM with persistence of opinion.

\subsubsection{Variance of average opinions}

\begin{figure}
\includegraphics[height=11cm,angle=270]{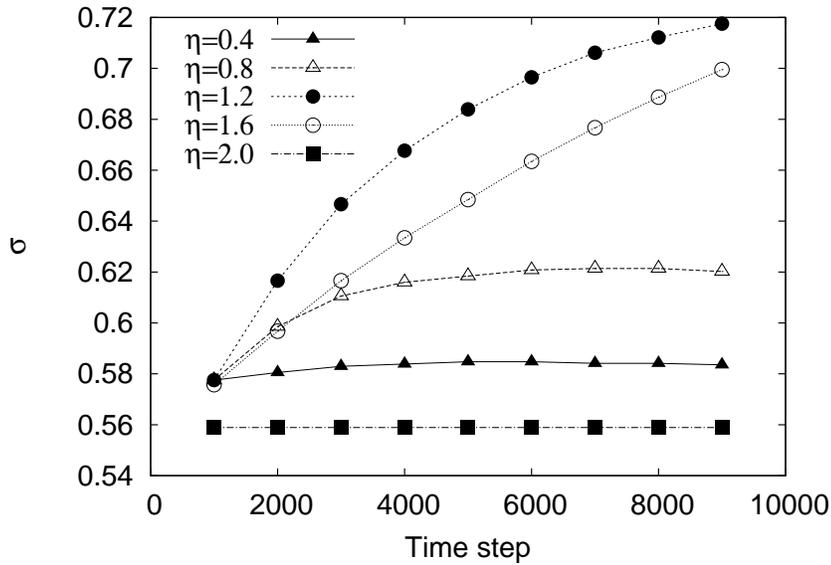}
\caption{Figure showing the evolution of the variance of average
opinions for various agreement parameters with timestep (errorbars are
within the points). $N=250$, $\alpha =
0$. LCM rules are used.}
\label{fig:six}
\end{figure}

In order to gain greater insight into the distribution of opinions, we
study the time evolution of the variance. Figure \ref{fig:six} shows
the evolution of the variance of average opinions for various
agreement parameters, clarifying what is seen in figure
\ref{fig:five}. The average variance typically reaches a plateau
within 10000 timesteps. The large timestep variances of opinions that
the models with $\eta =1.2$ and $\eta =1.6$ reach are very
similar. The variance drops dramatically above $\eta=1.2$. In the
simulation where $\eta>1.6$, neighbors pairs can only convince their
neighbors once they have a magnitude of opinion greater than
$O=0.8$. This explains the rapid initial increase in variance for
large $\eta$, which is the result of the extinction of agents with
centrist views. For $\eta \ge 2$ all interactions are via the
ambiguous rules - thus no LCM updates are permitted, and the variance
shows no change with time step.

\subsection{Comparisons with the binary USDF Model}

\label{sec:discreteresults}

The results presented in the previous section have shown that it is
necessary to make a further analysis of the absence of the 3 peak
structure in the density of long time scale average opinions (since
that is the structure in the binary model). To make this analysis, we
adjust our simulation to have only discrete opinions following
equation \ref{eqn:discrete}. The adjustments are minor. In the binary
case, we simply initialize with $O\in\{+1,-1\}$ and choose appropriate
model parameters so that no intermediate sized opinions are created
($\eta < \Delta O = 2/(M-1)$) where $\Delta O$ is the difference
between closest opinion states. For the binary simulations, $\Delta O
= 2$. We set $\alpha = 0$, since any memory factor would generate
non-binary opinions). We choose to simulate a set of agents with
$\eta=0.1$ following the LCM rules (for the binary simulation with
small $\eta$, LCM and ACM are identical, since the ambiguous rules are
never invoked). In order to investigate the long time step limit, we
analyze a population of $N=100$ agents with $T_{max}=25000$ time
steps. Those parameters were determined empirically to represent a
stable density of average opinions. We also have the opportunity to
examine systems with other numbers of opinions. A ternary model can be
formed by initializing $O\in\{+1,0,-1\}$ (or yes, no, unsure set)
where $\Delta O = 1$, but otherwise using the same parameters. Since
we use the LCM rules, agents that are `unsure' may not propagate their
opinions. We also simulate a 5 state model where $\Delta O = 0.5$
(strongly agree, agree, unsure, disagree, strongly disagree
set). Again, since the LCM rules are used, the `unsure' agents do not
propagate their opinions.

\begin{figure}
\includegraphics[height=62mm,angle=270]{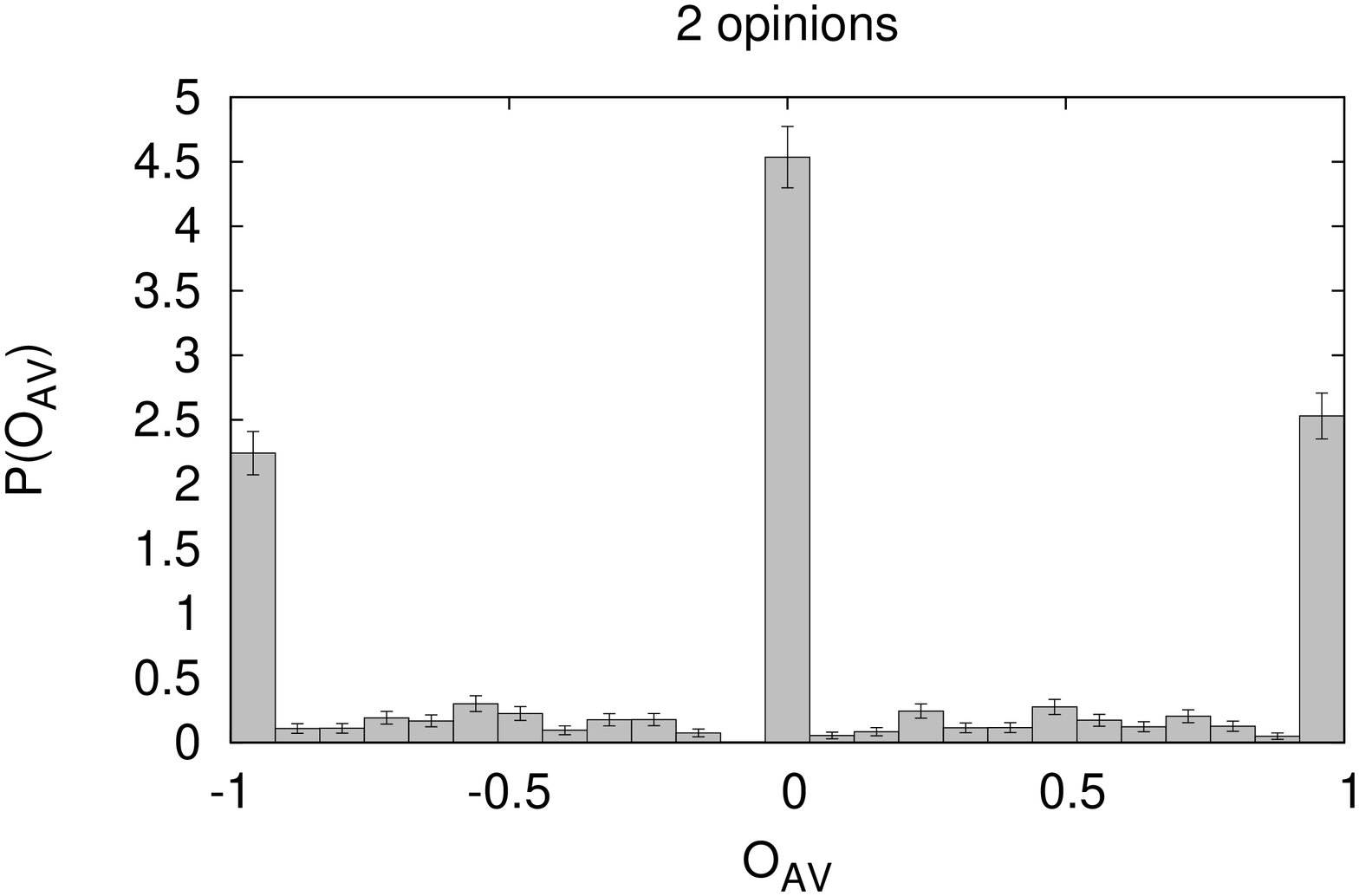}
\includegraphics[height=62mm,angle=270]{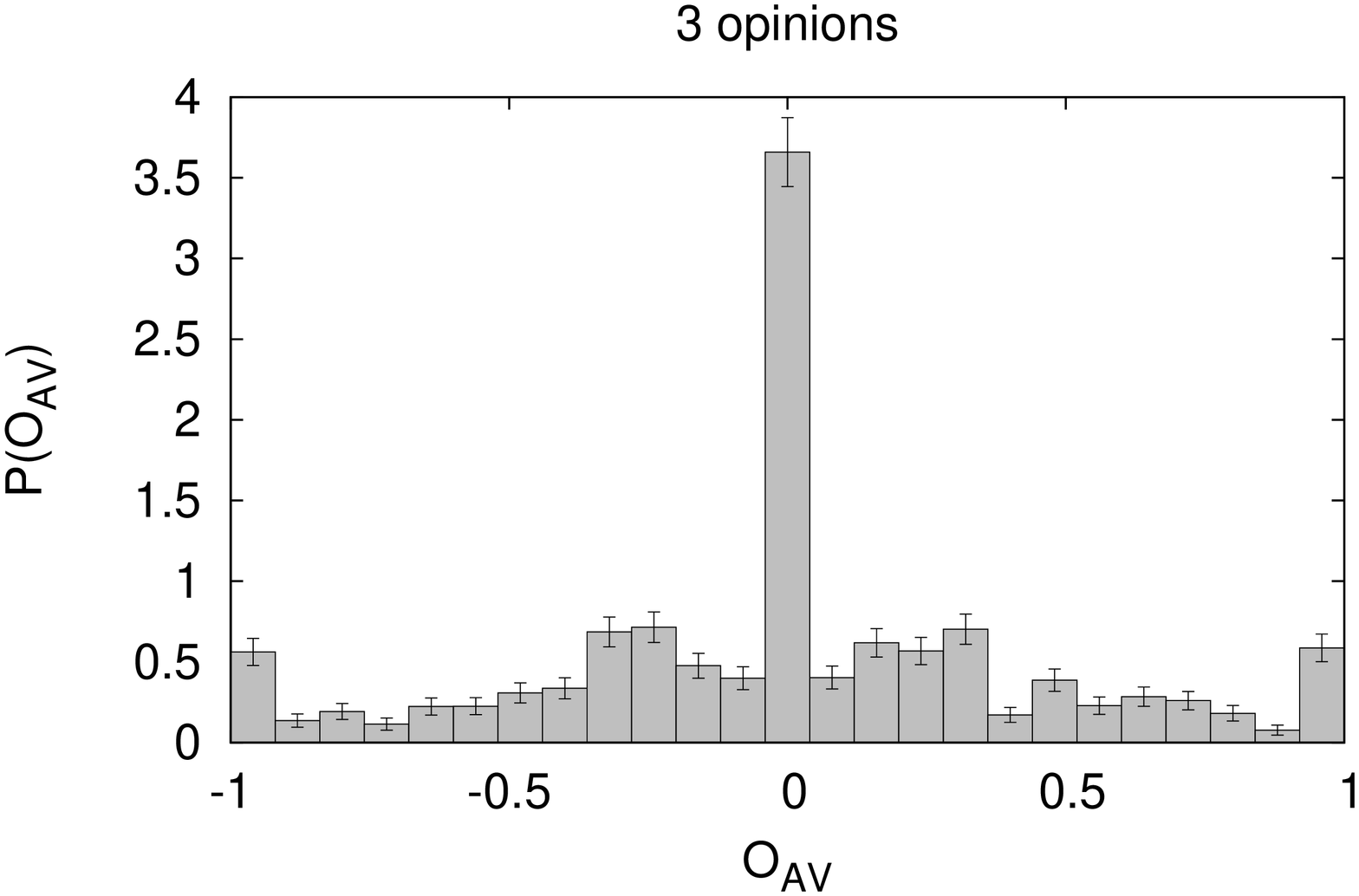}\\
\includegraphics[height=62mm,angle=270]{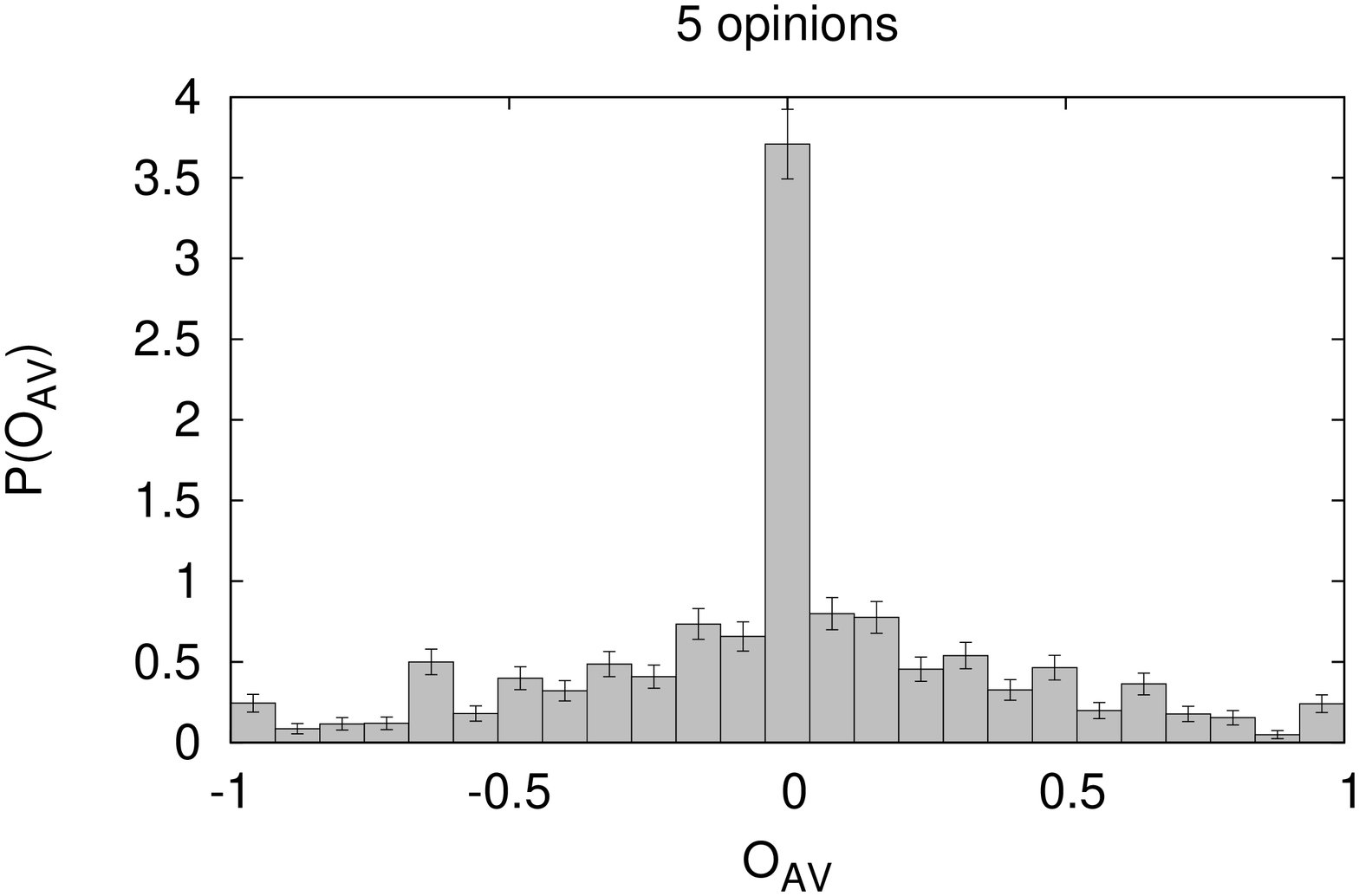}
\includegraphics[height=62mm,angle=270]{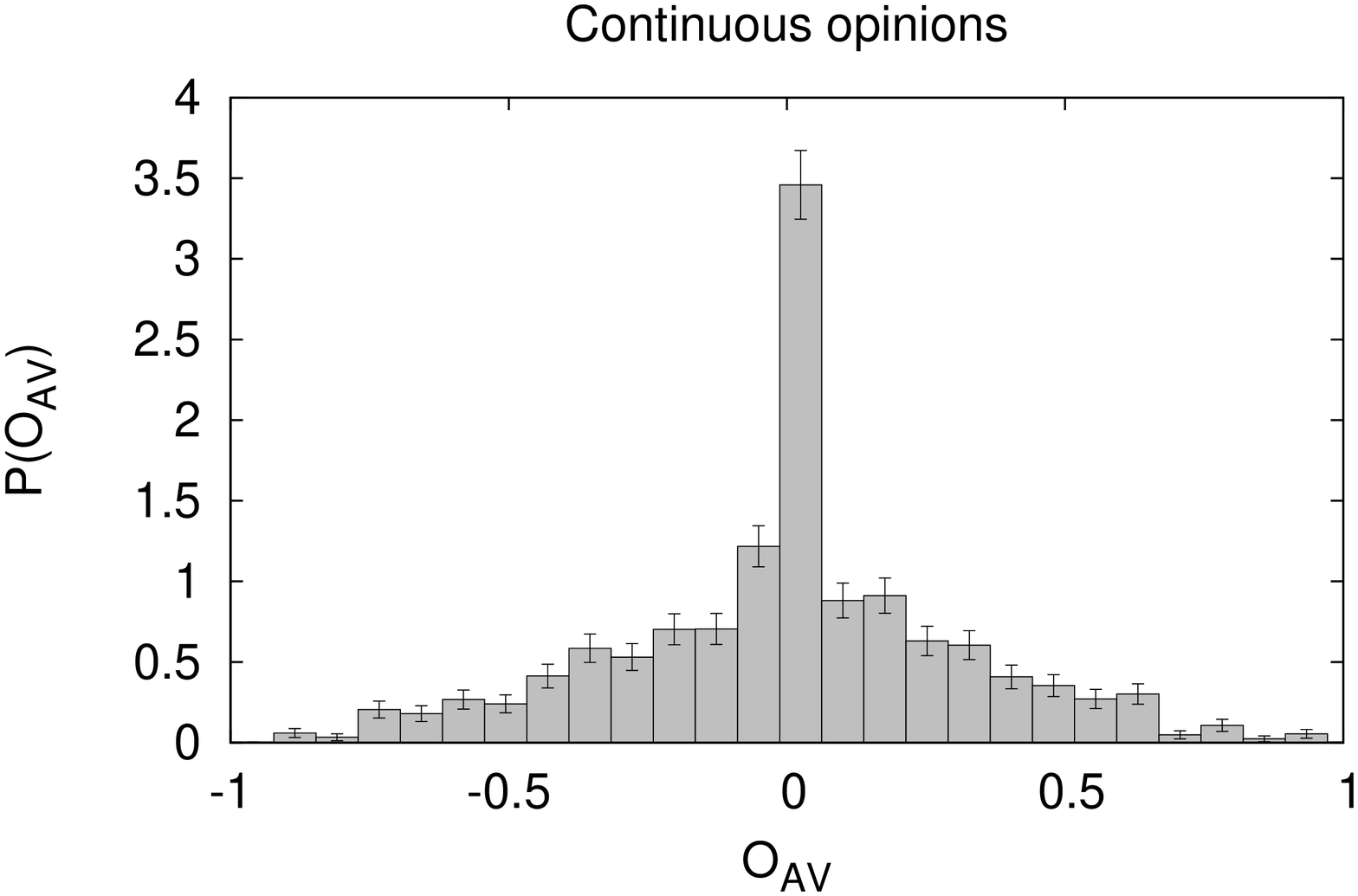}
\caption{Histograms showing the distribution of average opinions in an
ensemble of $N_{\rm ens} = 1000$ runs at time step $T = 25000$. The
plots are normalized so that the total area of the boxes is unity and
the graphs represent the density of average opinions. It shows the
changing of the binary USDF model into a continuous model. $\eta=0.1$,
$N_P=100$ and $\alpha=0$. The LCM rules were used.}
\label{fig:seven}
\end{figure}

\begin{figure}
\includegraphics[height=62mm,angle=270]{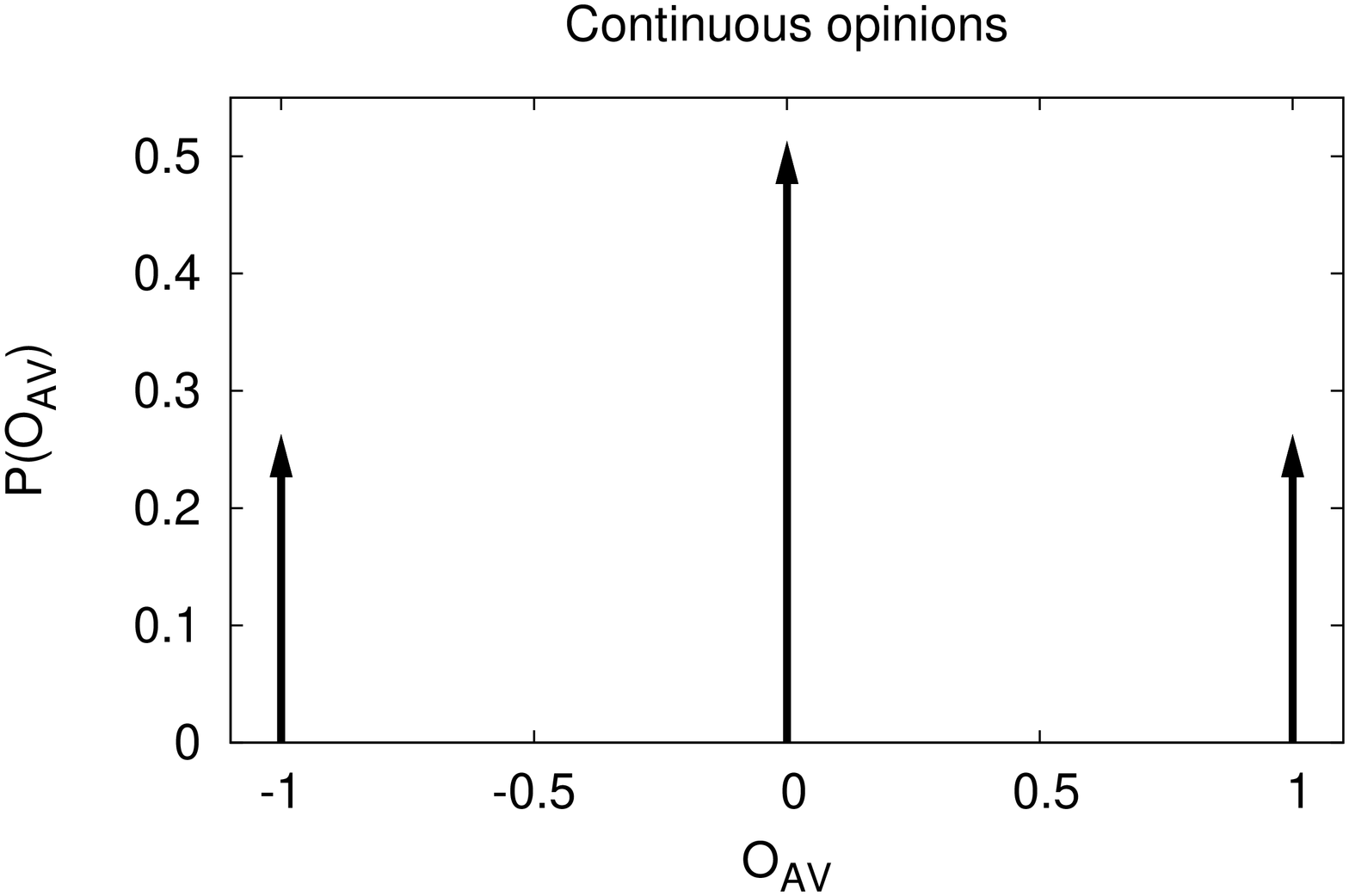}
\includegraphics[height=62mm,angle=270]{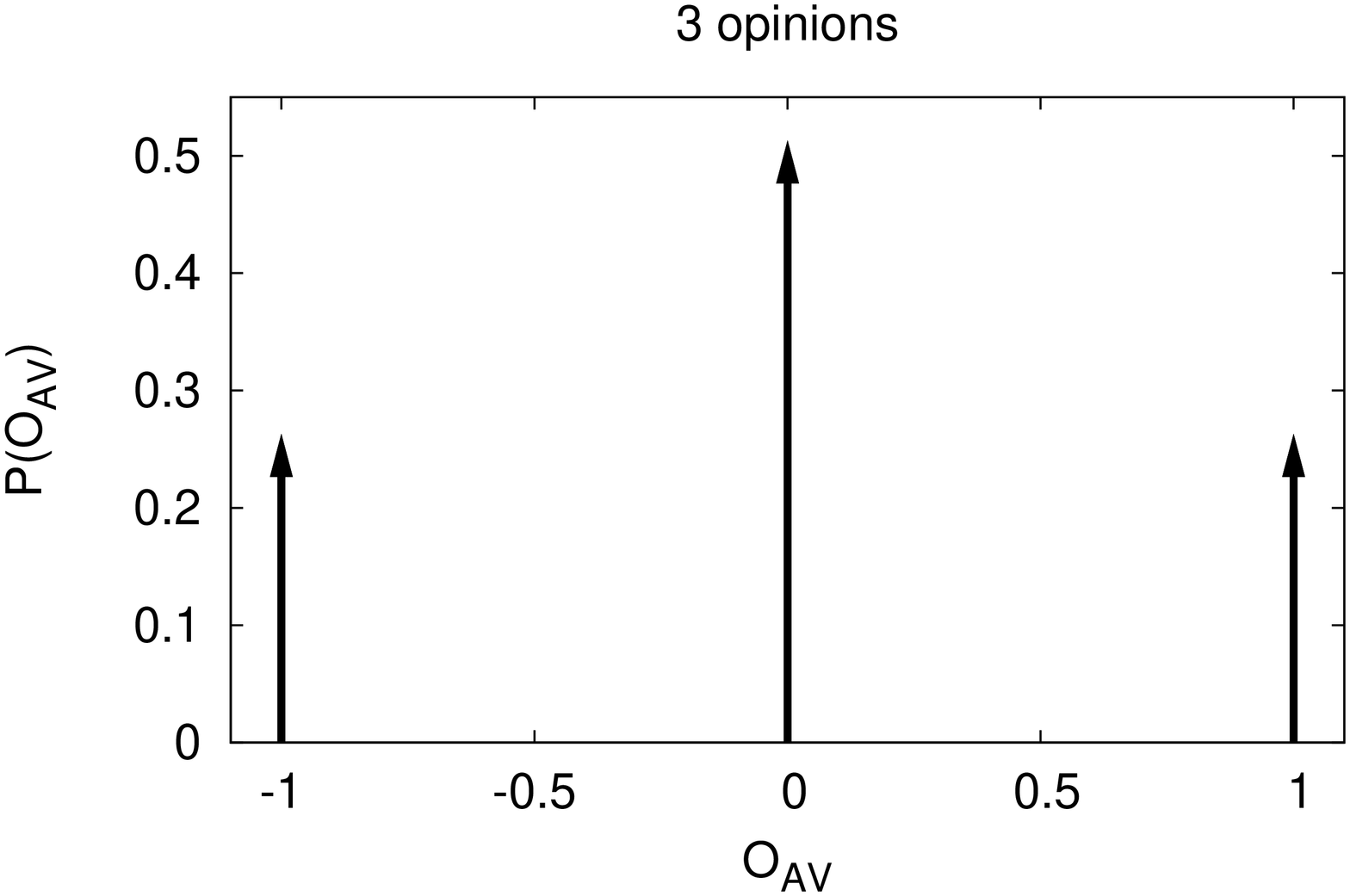}\\
\includegraphics[height=62mm,angle=270]{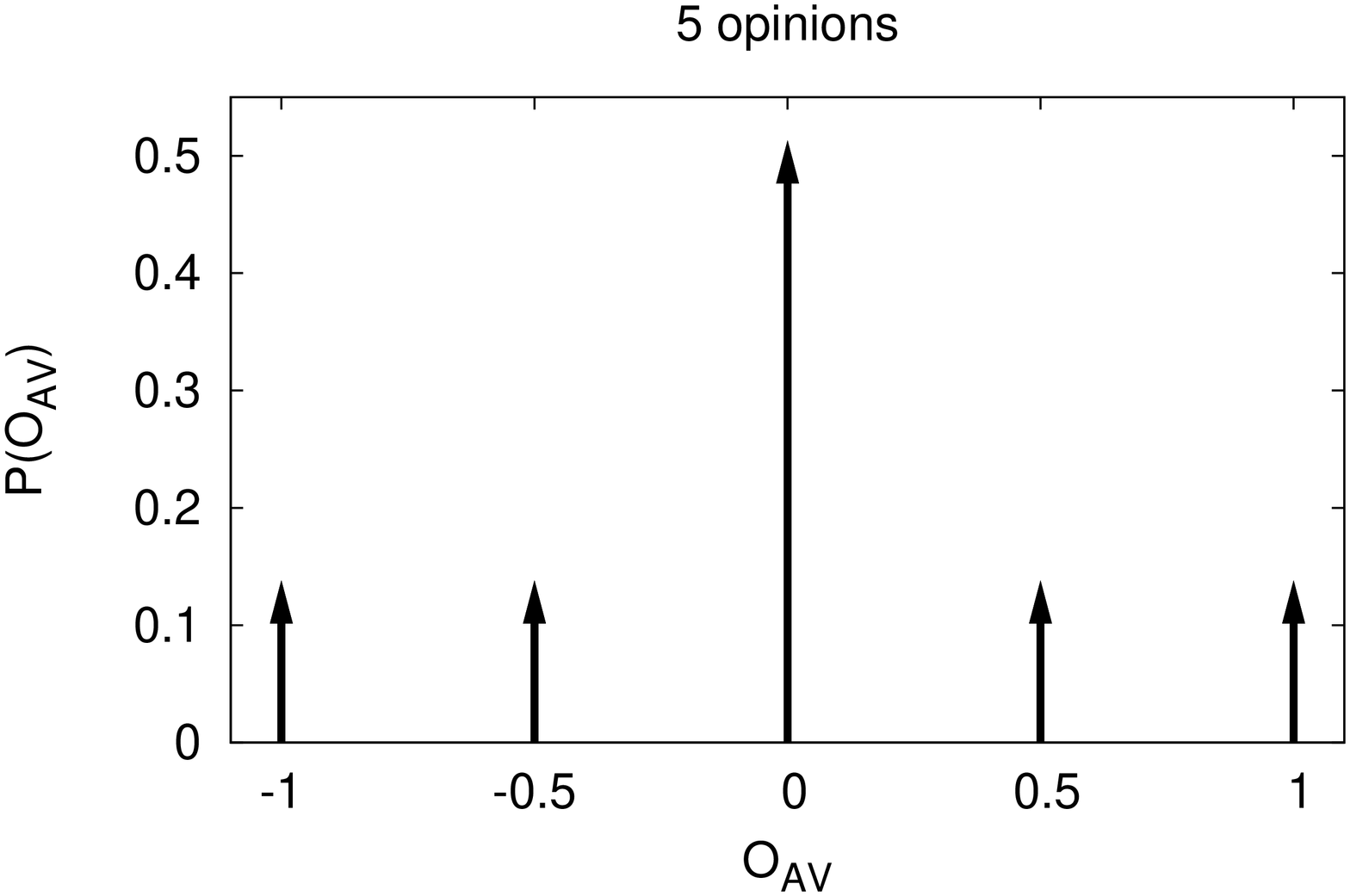}
\includegraphics[height=62mm,angle=270]{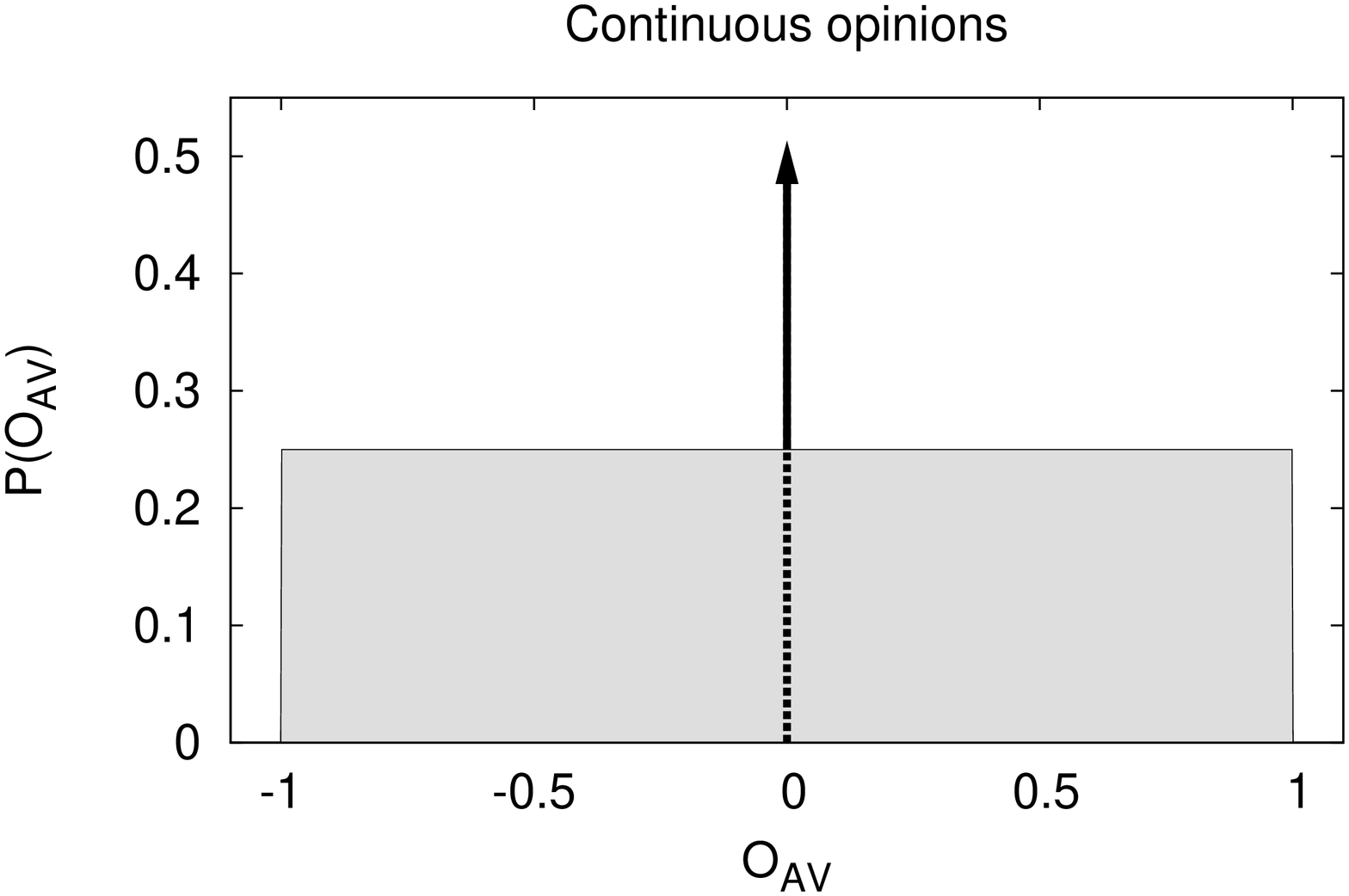}
\caption{Analytic long time results, following table
\ref{table:threeopinion}. Arrows represent $\delta$-functions with their weight represented by their height.}
\label{fig:eight}
\end{figure}

Figure \ref{fig:seven} was produced by binning the average opinion on
the final time step of each run in the ensemble. It shows the evolution of the
binary USDF model via 3 and 5 state models into a continuous model. Top left, 2 opinion, $M=2$
(Sznajd or binary USDF model), top right $M=3$, bottom left $M=5$ and
bottom right continuous opinions. Each simulation was carried out with
a population of $N=100$, $T_{max} = 25000$ loops over $N_{ens}=1000$
ensembles using the LCM rules with $\eta = 0.1$
and $\alpha = 0$.

There are significant differences in the density of average opinions
formed in the different discrete models, as seen in figure
\ref{fig:seven}. The binary USDF (Sznajd model) results (top left)
show that a three peaked system emerges in the long time limit. This
is expected as there are 3 possible stable outcomes to a binary system
in the long timescale limit, either all the agents take opinion A, all
the agents have opinion B or there is a perfect split, 50\% of the
agents taking each view on neighboring sublattices. This is a direct
consequence of the disagreement rule, which allows the space modulated
state with $O_{\rm av} = 0$ to form.

The same logic allows us to determine the steady states of a
three-opinion population of agents. It is useful to understand which
steady states we expect in the long time scale limit. The stable
states of a three opinion model can be easily determined and are shown
in the table (table \ref{table:threeopinion}). These are the states in
which application of any of the update rules to any pair in the
population leads to exactly the same configuration. Clearly the most
obvious steady states are where all the members of the population have
the same opinion. There are 3 of these in the ACM (all A, all B and
all C), and 2 in the LCM (all A and all C, since in the LCM the B
state becomes extinct as it is not able to propagate). In addition to
these states, states with modulated opinion are stable, with two types
of alternating A and C states (AC and CA). This distribution of
average opinions matches the results from simulations of the 3 opinion
model which we show in Fig. \ref{fig:seven} (top right).

\begin{table}
\tbl{Long time scale stable states of two, three and five state
models. Long time scale stable states of a five opinion model are
similar to those of the Sznajd model, but an average opinion of zero
is significantly more likely. For large number of states $M$, the
ratio of the magnitude of central (modulated) to outlying
(homogeneous) peaks is $1/M$. If there are no disagreement rules, modulated
states are not stable and the long timescale distribution is uniform.}
{\begin{tabular}{lll} 
\toprule 
Model and states & Steady state & Deg.\\\colrule 
Binary, & all A ($O_{\rm av} = 1$), all B ($O_{\rm av} = -1$) & 1\\  
A (=yes), B (=no) & alternating AB ($O_{\rm av} = 0$) & 2 \\\colrule
Ternary, & all A ($O_{\rm av} = 1$), all C ($O_{\rm av} = -1$) & \\  
A (=yes), B (=unsure), C (=no) & or all B (ACM only $O_{\rm av} = 0$) & 1\\
& alternating AC ($O_{\rm av} = 0$) & 2 \\\colrule
Five state, A (=strong agreement), & all A ($O_{\rm av} = 1$), all B ($O_{\rm av} = 1/2$), \\ 
B (=agreement), C (=unsure), & all C (ACM
only, $O_{\rm av} = 0$) & \\
D (=disagree),  & all D ($O_{\rm av} = -1/2$) or all E ($O_{\rm av} = -1$)& 1  \\
E (=strong disagreement) & alternating AE ($O_{\rm av} = 0$) or BD ($O_{\rm av} = 0$)& 2\\
\botrule
\end{tabular}}
\label{table:threeopinion}
\end{table}

We repeat the calculations for models with more opinion states. As can
be seen in Fig. \ref{fig:seven} (bottom left and right) for 5 or more
opinion levels, little difference is found between discrete and
continuous opinions. With a 5 state opinion base there is already a
very high probability of producing a central average opinion. Here, we
kept $\eta = 0.1$ constant. However, if the limit $M\rightarrow\infty$
was taken keeping $\Delta O < \eta$, the result would be a flat
continuous distribution with constant density $1/4$ (since half the
end states have all agents exhibiting the same opinion, $O_{\rm min} =
-1 < O_{\rm av} < O_{\rm max} = 1$ with weight $1/(O_{\rm max} -
O_{\rm min})$) and a $\delta$-function spike at $O_{\rm av} = 0$
represents end states with modulated opinion, i.e. $P(O_{\rm av}) = 1/4 +
\delta(O_{\rm av}) / 2$.

\section{Concluding remarks}

\label{sec:conclusions}

We have proposed an extension to the binary ``United we
stand, divided we fall'' opinion dynamics of Sznajd-Weron, which can handle continuous opinion distributions. 
We used the most symmetric extension
of the binary model in order to take account of the consequences of
disagreement (confrontation) within a population of agents in
conjunction with agreement (debate). We have also developed rules for
persistence of opinion (memory) and for the propagation of centrist
views. We carried out Monte Carlo simulations. Memory effects were
found to significantly modify the variance of average opinions in the
large timescale limits of the models. We described the breakdown of
binary opinion dynamics on approaching the continuum limit. We
compared Monte Carlo results with a consideration of long time-step steady states. Our main conclusion is that the outcomes of USDF models are strongly modified when agents are
permitted more than 3 opinions.

Geographically the lattice approach is incomplete as influences of
opinion are likely to form a complex network, combined with a local lattice
and also global influences (e.g. the media). There are a number of possible directions for further studies. For example it might be interesting to extend the model
to exist on the scale-free networks of Barbar\'{a}si \cite{barbarasi1999a}. It is estimated that there is a high degree of connectivity in real-world networks, much greater than in the simple
lattice where $z=2$ and the inter-agent distance is simply $d = |i-j|$ \cite{barbarasi1999a}. Choosing
to place agents in networks rather than on lattices may lead to a more detailed description of
opinion dynamics, although understanding long time scale behavior would be more difficult. Including external influences, non-conformists and global effects are all likely to improve insight. To finish, we would also like to point out that a simple propagation of average views of agents with similar magnitude opinions may not be sufficient. Sometimes the opinions of others are so extreme that they can polarize even centrists away from their views.

\section*{Acknowledgments}

JPH would like to acknowledge support from EPSRC grant no. EP/C518365/1.

\bibliographystyle{unsrt}
\bibliography{extended_sznajd}

\end{document}